\documentclass[times,review]{elsarticle}
\usepackage{framed,multirow}

\usepackage{amssymb}
\usepackage{latexsym}
\usepackage{hyperref}
\usepackage{wrapfig}

\usepackage{url}
\usepackage{xcolor}
\definecolor{newcolor}{rgb}{.8,.349,.1}

\usepackage{tikz}
\usepackage{amsmath}
\usepackage[linesnumbered,ruled,vlined]{algorithm2e}
\usepackage{subcaption}
\usepackage{ulem}
\usepackage{setspace}

\definecolor{dark_blue}{HTML}{2E3B8E}
\definecolor{gris_onix}{HTML}{ADAAA3}

\journal{Journal of Computational Physics}

\begin{document}

%\verso{A. Quiriny \textit{et al}}

\begin{frontmatter}

\title{X-Mesh: A new approach for the simulation of two-phase flow with sharp interface}

\author[1]{Antoine {Quiriny}}
\author[1]{Jonathan {Lambrechts}}
\author[2,3]{Nicolas {Moës}}
\author[1]{Jean-François {Remacle}}
\address[1]{Institute of Mechanics, Materials and Civil Engineering (iMMC), Avenue Georges Lemaître 4, 1348 Louvain-la-Neuve, Belgium}
\address[2]{Nantes Université, École Centrale de Nantes, rue de la Noë, 44321 Nantes, France}
\address[3]{Institut Universitaire de France (IUF)}

%\received{1 May 2013}
%\finalform{10 May 2013}
%\accepted{13 May 2013}
%\availableonline{15 May 2013}
%\communicated{S. Sarkar}

\begin{abstract}
%%%
Accurate modeling of moving boundaries and interfaces is a difficulty present in many situations of computational mechanics. We use the eXtreme Mesh deformation approach (X-Mesh) to simulate the interaction between two immiscible flows using the finite element method, while maintaining an accurate and sharp description of the interface without remeshing. In this new approach, the mesh is locally deformed to conform to the interface at all times, which can result in degenerated elements. The surface tension between the two fluids is added by imposing the pressure jump condition at the interface, which, when combined with the X-Mesh framework, allows us to have an exactly sharp interface. If a numerical scheme fails to properly balance surface tension and pressure gradients, it leads to numerical artefacts called spurious or parasitic currents. The method presented here is well balanced and reduces such currents down to the level of machine precision.
%%%%
\end{abstract}

\end{frontmatter}

%\begin{keyword}
%% Keywords
%\KWD Keyword1\sep Keyword2\sep Keyword3
%\end{keyword}

%\linenumbers


\section{Introduction}
The finite element method has been predominant in computational
mechanics since the early 1970s. Besides the fact that finite elements are based on a robust mathematical theory \cite{brenner2008mathematical}, a fundamental reason why finite elements are popular is the fact that it can use unstructured meshes
that allows accurate discretization of boundaries or interfaces for example between two different materials. When these interfaces are known in advance, they can be represented in the digital blueprints of the parts to be modeled. Modern automatic mesh generators such as Gmsh \cite{geuzaine2009gmsh} can generate meshes that are conforming to these interfaces. When a physical model involves discontinuous material properties, interfaces can develop/nucleate, grow, change topology or disappear. These interfaces can be material or immaterial, depending on whether the material particles move with the interface or not. The position and speed of these interfaces is not known in advance and is part of the computation.

In this paper, we study the case of material interfaces between two immiscible fluids whose material properties such as viscosity or density are discontinuous. The existence of surface tension adds for the possibility of a pressure jump at the interface between the two fluids. There is an extensive literature on the use of finite elements to simulate two-phase flows \cite{prosperetti2009computational, FRACHON201977, NAGRATH20054565}. One can classify the approaches in two main categories: interface-tracking and interface-capturing
methods.

In the case of interface tracking, the mesh has the duty
to track the interface that remains consistently meshed
during its movement. The most accurate methods in this category are the ALE methods \cite{hughes1981lagrangian} where the nodes of the mesh are moved with the velocity of the interface. ALE methods have excellent conservation properties \cite{lesoinne1996geometric}. They are accurate and relatively simple to implement, but their biggest drawback is that they do not allow -- at least in their pure version -- large deformations of the interface or topology changes. Some papers \cite{alauzet20073d} propose to use mesh adaptation in combination with ALE but frequent mesh adaptations have the consequence to introduce time discontinuities in the solution. It is possible to mitigate this issue by only performing few mesh adaptations and use a fixed point algorithm \cite{alauzet2003transient}.  Nevertheless, these techniques move away from the original simplicity of ALE methods by introducing the complexity and relative fragility of a mesh adaptation in time.

Interface capturing methods generally consider a fixed mesh and an indicator function, discretized on this fixed mesh, which indicates the position of the interface \cite{PILLIOD2004465, HIRT1981201}. For example, in level set methods \cite{marchandise2007stabilized}, the indicator function is the signed distance to the interface and the iso-zero of this function represents the interface. Interface capturing methods have complementary advantages and disadvantages to interface tracking methods. They allow topology changes but require profound changes in finite element formulations.

It should be noted that there are sharp and diffuse versions of interface tracking and capturing methods.  Diffuse methods consist in regularizing the physical properties of the two fluids, i.e. in smoothing the viscosities and densities of the two fluids over a $\epsilon$ thickness. It can be shown that the error committed by diffusing the physical properties is of the order of $\sqrt{\epsilon}$ \cite{azaiez2016two}  which means
that diffuse methods can only be accurate if the mesh is refined -- possibly anisotropically -- in the vicinity of the interface.

Recently, we developed a tracking method that comply with the following X-Mesh specifications: i) to be sharp, ii) to work on a fixed topology mesh
and iii) to allow large topological changes of the interfaces. The first X-Mesh paper was dealing with the immaterial interface computation of phase-changes using
 the Stefan model \cite{X-Mesh}. This paper aims to develop X-Mesh in the context of material interfaces and in particular to solve two-phase flow problems. \\
 
    The key idea of X-Mesh is to allow elements to deform up to zero measure. For example, a triangle can deform to an edge or even a point. This idea is rather extreme and totally revisits the interaction between the meshing community and the computational community who, for decades, have striven to interact through beautiful meshes. Considering zero-measure elements allows the mesh to deform in a time continuous manner while providing: i) Relaying: The interface is transferred from one node to another node located at the same position giving interface propagation, ii) seeding: An element is reduced to a point that nucleates the interface and then deploys (and later relays the front to outer nodes) and iii) Annihilation: When two interfaces contact each other, there is the possibility to detach the interface from the nodes to model interface coalescence.
 \section{Governing Equation}
We consider the flow of two immiscible, incompressible and Newtonian fluids interacting together via an interface $\Gamma$. The evolution of the velocity and pressure of each fluid in time is described by the incompressible Navier-Stokes equations:
\begin{align} \label{NS}
\rho \left(\partial_{t} \mathbf{u}+\mathbf{u} \cdot \nabla \mathbf{u} \right) &=-\nabla p+\mu \nabla^{2} \mathbf{u}+\mathbf{f} \\ \nonumber
\nabla \cdot \mathbf{u} &=0
\end{align}
where $\mathbf{u}$ is the velocity field $(u,v)^T$, $p$ is the pressure, $\rho$ is the density, $\mathbf{f}$ the forces at distance (gravity) and $\mu$ the dynamic viscosity.\\
\iffalse
\begin{tikzpicture}
    \useasboundingbox (-40mm,-152.5mm) rectangle (40mm,-72.5mm);
    
    \filldraw[fill=gris_onix] (-40mm,-152.5mm) rectangle (40mm,-72.5mm);
    \filldraw[fill=dark_blue] (10mm,-90mm) circle (10mm);
    \filldraw[fill=dark_blue, domain=-1.27*pi:1.27*pi, variable=\x]
      (-40mm, -152.5mm) 
      -- plot (\x,{0.3*sin(\x r)-12.5})
      -- (40mm, -152.5mm)
      -- cycle;
      
    \draw [domain=-1.27*pi:1.27*pi] plot (\x,{0.3*sin(\x r)-12.5});
    \node[text width=4cm] at (40mm,-119mm) {$\Gamma$};
    \node[text width=4cm] at (38mm,-80mm) {$\Gamma$};
    \node[text width=4cm, color=white] at (26mm,-95mm) {$\Omega_1$};
    \node[text width=4cm] at (-10mm,-110mm) {$\Omega_2$};
    \node[text width=4cm, color=white] at (-10mm,-140mm) {$\Omega_1$};
\end{tikzpicture}
\fi
\begin{figure}[!ht] 
\centering
\includegraphics[width=0.6\textwidth]{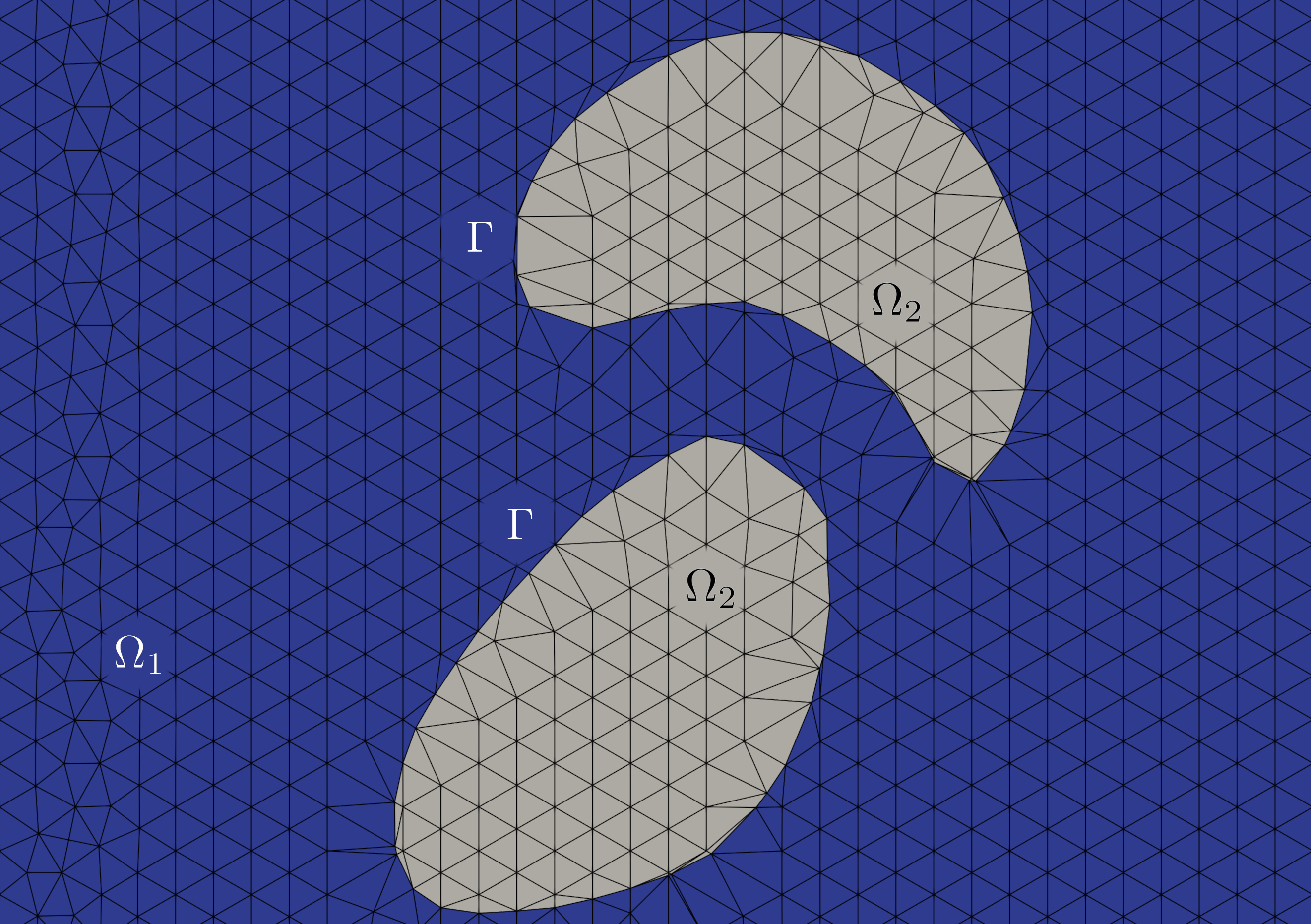}
\caption{Two-phase flow sketch.}
\label{fig:tpf}
\end{figure} \\
The two fluid phases are noted $\Omega_1$ and $\Omega_2$ (see Figure \ref{fig:tpf}) and have different dynamic viscosity and density $(\rho_1, \mu_1)$ and $(\rho_2,\mu_2)$, respectively. These material properties are thus discontinuous across the interface $\Gamma$. The Navier-Stokes equations are solved simultaneously on both subdomains and boundary conditions at the interface are necessary. The two equations \ref{NS} correspond to the conservation of momentum and the conservation of mass. These conservation laws must be respected at the interface between the two fluids. The conservation of mass leads to the condition of continuous velocity in the normal direction at the interface, thus preventing the transfer of mass between the two phases:
\begin{align*}
    [\mathbf{u}] \cdot \mathbf{n} = \mathbf{0}
\end{align*}
where the brackets $[ \cdot ]$ refer to the jump at the interface.\\
In addition to the condition of impermeability of the interface, the conservation of the momentum in the perpendicular direction to the interface must be respected. To satisfy this condition, the jump in normal stress is counterbalanced by the surface tension
\begin{align} \label{qdm}
    \left[ -pI + \mu \hspace{2pt} \frac{1}{2} \left( \nabla \mathbf{u} + \nabla \mathbf{u}^T \right) \right] \cdot \mathbf{n} = \sigma \kappa \mathbf{n}
\end{align}
with $I$ the identity operator, $\sigma$ the surface tension coefficient and $\kappa$ the curvature of the interface.

In two-phase flows, the main challenge is to compute the position of the interface accurately.
When we talk about an accurate interface position, we mean two things. We must be able to compute complex
mobile interfaces whose topology changes but we should ideally compute an interface
whose position is such that the mass of each of the two fluids is conserved.
In this first X-Mesh multi-phase paper, we focus on the first of these two challenges and we will show that it is possible to track very
complex fronts with topology changes using a mesh with a fixed connectivity.  Mass conservation is not ensured in the method presented in this paper and will be discussed shortly in the conclusion, however it's an issue we would like to address in future work.
To describe the position of the interface, we use here a classical technique. An indicator function $\phi(x,y)$
is discretized on the same mesh as the one used to discretize the fluid pressure and velocity.
This \emph{level set} function \cite{sethian2003level} classically represents the signed distance to $\Gamma$.
Thus, in the standard setting, the evolution in time of the level set function $\phi$
and therefore of the interface is governed by the advection equation:
\begin{align} \label{levelset}
    \partial_t \phi + \mathbf{u} \cdot \nabla \phi = 0
\end{align}
with $\mathbf{u}$ corresponding to the velocity of the fluids obtained in equations \ref{NS}.
\section{Finite Element Solver}
In this section we describe our numerical method for the resolution of two-phase flows. We choose to work with the finite element method and more specifically with the Galerkin approach and stabilized P1 elements for solving both the level set equation \ref{levelset} and the Navier-Stokes equations \ref{NS}. The algorithm to deform the mesh in order to follow the interface is then explained in section \ref{xmesh}. Finally, a sequential coupling between these steps is presented.

\subsection{Navier-Stokes solver}
The different flows targeted by our approach can be dominated by advection, the continuous Galerkin finite element method must then be stabilized to avoid spurious oscillations. Several stabilizations have been developed for this purpose. In this work, we use the Streamline Upwind/Petrov-Galerkin (SUPG) method \cite{brooks1982streamline}. In order to minimize the size of the linear system to solve, our unknowns are positioned at the nodes of our mesh. However, this does not respect the Babuska-Brezzi condition \cite{brezzi1974existence} and leads to the appearance of high frequencies in the pressure field. We get around this problem by using the popular Pressure-Stabilizing/Petrov-Galerkin (PSPG) stabilization \cite{hughes1986new}.

In our X-Mesh approach, we use a mesh that has a fixed connectivity: only vertex positions change with time. Coordinates of the mesh vertices are thus a variable denoted $\mathbf{x}(t)$.
By deforming the mesh to match the interface, the mesh is neither fixed nor moving in a Lagrangian manner. We use the Arbitrary Lagrangian Eulerian (ALE) method to account for the mesh velocity in the Navier-Stokes equations. As shown in equation (4), we consider a non-conservative ALE formulation where the integration is performed on the same mesh configuration and the mesh velocity $\mathbf{u}_{mesh}$ is subtracted from the advection velocity.
The time is discretized using a constant time step $\Delta t$. We denote discrete time instants of variables that depend on time -- mesh positions $\mathbf{x}(t)$ for example --  as $\mathbf{x}_n = \mathbf{x}(n \Delta t)$.
The mesh velocity is thus $\mathbf{u}_{mesh} =(\mathbf{x}_{n} - \mathbf{x}_{n-1})/\Delta t$. For the temporal integration we use the implicit Euler scheme and we obtain the discrete formulation with the finite element method.

Consider $\mathbf{S}_{\mathbf{u}}$ and $\mathbf{S}_{p}$, the solution spaces of $\mathbf{u}_{n+1}$ and $p_{n+1}$, respectively and their test functions $\left(\mathbf{v}_{n+1}, q_{n+1}\right) \in \mathbf{V}_{\mathbf{u}} \times \mathbf{V}_{p}$.\\

The fully discrete formulation of \eqref{NS} is to find $\left(\mathbf{u}_{n+1}, p_{n+1}\right) \in \mathbf{S}_{\mathbf{u}} \times \mathbf{S}_{p}$ such that for any $\left(\mathbf{v}_{n+1}, q_{n+1}\right) \in \mathbf{V}_{\mathbf{u}} \times \mathbf{V}_{p}$:
\begin{align} \nonumber
     \int_{\Omega_n} \rho \hspace{2pt} \left( \left(\mathbf{u}_{n+1} - \mathbf{u}_{n} \right) {\frac{1}{  \Delta t}} + \mathbf{u}_{n+1} \cdot \nabla \mathbf{u}_{n+1} -\mathbf{u}_{mesh}\cdot \nabla \mathbf{u}_n \right) \cdot \mathbf{v}_{n+1} 
    d \Omega \\ \label{NS_discr} + \int_{\Omega_n} \mu \nabla \mathbf{u}_{n+1} : \nabla \mathbf{v}_{n+1}   d \Omega - \int_{\Omega_n}  \nabla p_{n+1} \cdot \mathbf{v}_{n+1}  d \Omega  =   \int_{\Omega_n} \rho \hspace{2pt} \mathbf{g}_{n+1} \cdot \mathbf{v}_{n+1} d \Omega  + \text{SUPG}
\end{align}
\begin{equation*}
 \int_{\Omega_n}\nabla \cdot \mathbf{u}_{n+1} \hspace{1.5pt}  q_{n+1} d \Omega + \text{PSPG}  =0 
\end{equation*}
With $\text{SUPG}$ the term for the SUPG stabilization and $\text{PSPG}$ the term from the PSPG stabilization that can be expressed in function of the residual $\mathcal{R}$ of the equation:
\begin{align*}
    \text{SUPG} &= \int_{\Omega_n} \tau_{\text{SUPG}}  \mathcal{R}_{n+1} \left( \mathbf{u}_{n+1} - \mathbf{u}_{mesh} \right) \cdot \nabla \mathbf{v}_{n+1}  d\Omega\\
    \text{PSPG} &= \int_{\Omega_n} \tau_{\text{PSPG}} \nabla q_{n+1} \mathcal{R}_{n+1} d\Omega\\
    \mathcal{R}_{n+1} &=  \left( \rho \hspace{2pt} \left(\mathbf{u}_{n+1}  -\mathbf{u}_{n} \right) {\frac{1}{  \Delta t}} + \rho \hspace{2pt} \left(\left(\mathbf{u}_{n+1} - \mathbf{u}_{mesh} \right) \cdot \nabla \mathbf{u}_{n+1}\right) - \nabla p_{n+1} - \rho \hspace{2pt} \mathbf{g}_{n+1} \right) 
\end{align*}
with $\tau_{\text{SUPG}}$ and $\tau_{\text{PSPG}}$ the coefficient for the SUPG and PSPG stabilisation respectively. 

The classical finite element spatial discretization yields a nonlinear system of equations for $\mathbf{u}_{n+1}$ and $p_{n+1}$ that can be solved using a Newton scheme.
For the Navier-Stokes solver we used the open-source software Migflow \cite{constant2019implementation}. 
\subsection{Level set solver}
The position of the interface between the two fluids is described by the iso-contour $\phi=0$ of the level set function and its temporal evolution is determined by the avection equation \ref{levelset}. As for the resolution of the Navier-Stokes equations, we stabilize the advection via an SUPG term \cite{brooks1982streamline}. The resolution of the equation is done on a fixed mesh, we thus drop the index notation for the computational domain $\Omega$ and the time integration is solved by a Cranck Nicolson method. Let $\mathbf{S}_{\phi}$ be the solution space of $\phi_{n+1}$ and $w_{n+1} \in \mathbf{V_{\phi}}$ be the test function associated to $\phi_{n+1}$. The discrete formulation can then be written as follows:\\

Find $\phi_{n+1} \in \mathbf{S}_{\phi}$ such that for any $\psi \in \mathbf{V}_{\phi}$:
\begin{align*}
    &\int_{\Omega} \left(\phi_{n+1}\hspace{1.5pt} \psi  -  \phi_{n} \hspace{1.5pt} \psi \right) {\frac{1}{\Delta t}} d\Omega   + \frac{1}{2}  \int_{\Omega}  \left(\mathbf{u}  \cdot \nabla \phi_n \hspace{1.5pt} \psi + \phi_n \nabla \cdot  \mathbf{u} \hspace{1.5pt} \psi \right) d\Omega  \\& + \frac{1}{2}  \int_{\Omega} \left( \mathbf{u} \cdot \nabla \phi_{n+1} \hspace{1.5pt} \psi  +  \phi_{n+1} \nabla \cdot  \mathbf{u} \hspace{1.5pt} \psi \right) d\Omega + \text{SUPG} = 0
\end{align*}
The SUPG stabilisation term depends of the residual of the advection equation and is given by:
\begin{align*}
    \text{SUPG} &= \int_{\Omega} \tau_{\text{SUPG}} \hspace{1.5pt} \mathbf{u} \cdot \nabla \mathbf{\psi} \hspace{1.5pt} \mathcal{R} d\Omega \\
    \mathcal{R} &= \left( \left(\phi_{n+1} - \phi_{n} \right) {\frac{1}{\Delta t}} + \frac{1}{2}  \left(\mathbf{u}  \cdot \nabla \phi_n + \phi_n \nabla \cdot  \mathbf{u} \right) + \frac{1}{2} \left( \mathbf{u} \cdot \nabla \phi_{n+1} +  \phi_{n+1} \nabla \cdot  \mathbf{u} \right) \right) \hspace{1.5pt} \psi
\end{align*}
The finite element method then allows us to spatially discretize the domain and we obtain a system of linear equations. Since we don't take into account the mesh displacement in the resolution of the advection equation, after moving the mesh with the deforming algorithm presented in the following section the value of the level set at the nodes aren't correct. To compute the new values at the nodes of the level set we use a fast-marching algorithm \cite{sethian1996fast} that computes an approximate signed distance to the front. Since the interface is completely embedded by the edges of the mesh, the seeds of the algorithm from which the level set value propagates in the mesh are the front nodes showed in Figure \ref{fig:xmesh6} (e). 

\subsection{Extreme mesh deformation -- Front relaying} \label{xmesh}
The idea behind the X-Mesh approach \cite{X-Mesh} is to deform a mesh with continuous node movements to constantly match the interfaces of interest, even in the case of topological changes of the fluids domains. The mesh however keeps a fixed topology. To achieve this goal, the approach allows elements to become degenerated, meaning that a triangle can deform down to an edge or even a point. This enables the mesh to deform continuously in time and ensure the relay of the front. The interface is transferred from one node to another located at the same position allowing the interface to propagate like a baton in a relay race.

There are two ways for a triangle to degenerate into an edge. In the first case, one of the three edges of the triangle collapses to a single point so the opposite angle to this edge degenerates to $0$ radians. Such an element is called \emph{a needle}. In the second case, one of the three vertices of the triangle tends to a point that belongs to its opposite edge so the angle associated with this node tends to $\pi$ radians. Such an element is called \emph{a cap}.

The use of degenerate or quasi-degenerate elements leads to two specific difficulties: \emph{conditioning and stability}. 

\subsubsection{Conditioning}

\begin{wrapfigure}{L}{0.32\textwidth}
\begin{center}
  \begin{tikzpicture}
\coordinate (i) at (0,0);
\coordinate (j) at (0:2);
\coordinate (k) at (60:1.5);
\node at (i) {$\bullet$};
\node at (j) {$\bullet$};
\node at (k) {$\bullet$};
\draw (i)--(j)--(k)--cycle;
\node at (-.2,0) {$i$};
\node at (0:2.2) {$j$};
\node at (62:1.75) {$k$};
\node[blue] at (30:.35) {$\theta_i$};
\node[blue] at (50:1.25) {$\theta_k$};
\node[blue] at (7:1.6) {$\theta_j$};
\end{tikzpicture}
\end{center}
  \caption{A triangle with its three internal angles $\theta_i$, $\theta_j$ and $\theta_k$.} 
  \label{fig:triangle}
\end{wrapfigure}
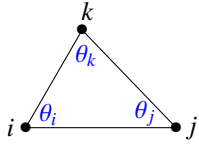

The first problem that appears with the use of degenerate or quasi-degenerate elements is the bad conditioning of the finite element matrices.
Assume a triangle with 3 vertices $i$, $j$ and $k$ and with internal angles $\theta_i$, $\theta_j$ and $\theta_k$ (see Figure \ref{fig:triangle}).   

\noindent The local stiffness matrix $[K]$ that corresponds to the discretization of the $-\nabla^2$ operator can be written in its cotangent form as
$${\small
    [K]
= 
    \begin{bmatrix}
    \cot(\theta_j)+\cot(\theta_k)  & -\cot(\theta_k) & -\cot(\theta_j) \\ 
    -\cot(\theta_k) &\cot(\theta_i)+\cot(\theta_k) &  -\cot(\theta_i) \\ 
    -\cot(\theta_k)&  -\cot(\theta_i) &  \cot(\theta_i)+\cot(\theta_j)  
\end{bmatrix}}
$$
Whether we are talking about a needle or an cap, at least one of the angles $\theta_{i,j,k}$ of a degenerate triangle will
tend to zero which implies that the cotangent of this angle will tend to infinity.
More precisely, the stiffness matrix $[K]$ necessarily contains a zero eigenvalue $\lambda_1 = 0$ which corresponds to the rigid body or constant
mode $v_1 = (1,1,1)$. When this element is degenerated, another eigenvalue $\lambda_2$ also tends to zero and the third one, $\lambda_3$  tends to infinity.

Let us imagine the case of a needle: vertex $i$ and vertex $k$ eventually coincide. The mode that correspond to zero eigenvalue $\lambda_2$ is
$v_2 = (1,0,-1)$. Mode $v_2$ is the only mode besides $v_1$ -- the rigid body mode -- that can exist in the finite element solution because
any other combination that involves $v_3$ leads to an infinite energy. Thus, at the zero measure limit, the finite element solution is such that
the nodal values of the unknown at nodes $i$ and $k$ are the same.

Now let's build  a cap: vertex $k$ moves at position
$${\bf x}_k = {\bf x}_i \alpha + {\bf x}_k (1-\alpha)~~,~~\alpha \in ]0,1[.$$
The mode that correspond to zero eigenvalue $\lambda_2$ is
$v_2 = (\alpha,1-\alpha,-1)$ which implies that in the zero measure limit, the finite element solution will be
linear along the straight line $ikj$.

Why was the difference made between a needle and a cap?
In both cases, the global stiffness matrix will suffer from poor conditioning but linear solvers -- even iterative ones --
behave very well in the case of very large eigenvalues. Any preconditioner will solve this problem.
In this paper, we proceed in the same fashion as in \cite{X-Mesh}. We simply limit the maximum value of
$\lambda_3$ by limiting the minimum value of the area of degenerated triangles. In future work, we will propose
robust preconditioners for X-Mesh.
The difference between a needle and a cap is in the stability as we will detail below. A needle imposes a local
constraint between two degrees of freedom while a cap couples the three degrees of freedom of the triangle. 

\subsubsection{Stability}

If stability is lost, then the convergence of the finite element method is lost as well. Historically, the finite element community took from the seminal paper of Babu{\v{s}}ka and Aziz \cite{babuvska1976angle}  that, to ensure finite element convergence, it was sufficient to generate meshes whose triangles
did not have very obtuse angles. Stability issues are thus essentially related to the presence of caps in the mesh.
Many have considered that this \emph{angle condition} was a necessary condition for the convergence of finite elements, but this is not correct: it is a sufficient condition -- Babu{\v{s}}ka and Aziz have never said the contrary. The angle condition can be significantly weakened
\cite{kuvcera2016necessary}. An isolated cap in a mesh will not cause any stability concerns. It is only when caps are joined together
in long bands that stability problems arise.  If usual P1 triangles are used, it is easy to show that a band of $n$ caps of individual length $h$ ($h = \|{\bf x}_j - {\bf x}_i\|$ in Figure \ref{fig:triangle}) prevents non-linear variations of the solution along the whole band length $nh$ (locking) and thus degrades the convergence from $h^2$ to $(nh)^2$.
On the other hand, as we have shown above, needles impose local constraints on the solution and these constraints therefore
do not propagate to great distances. We can therefore have as many needles as we want without worrying about stability.

\begin{figure}[!htb]
    \captionsetup{font=normal}
    \centering
    \begin{subfigure}[b]{0.32\textwidth}
        \centering
        \includegraphics[width=\textwidth]{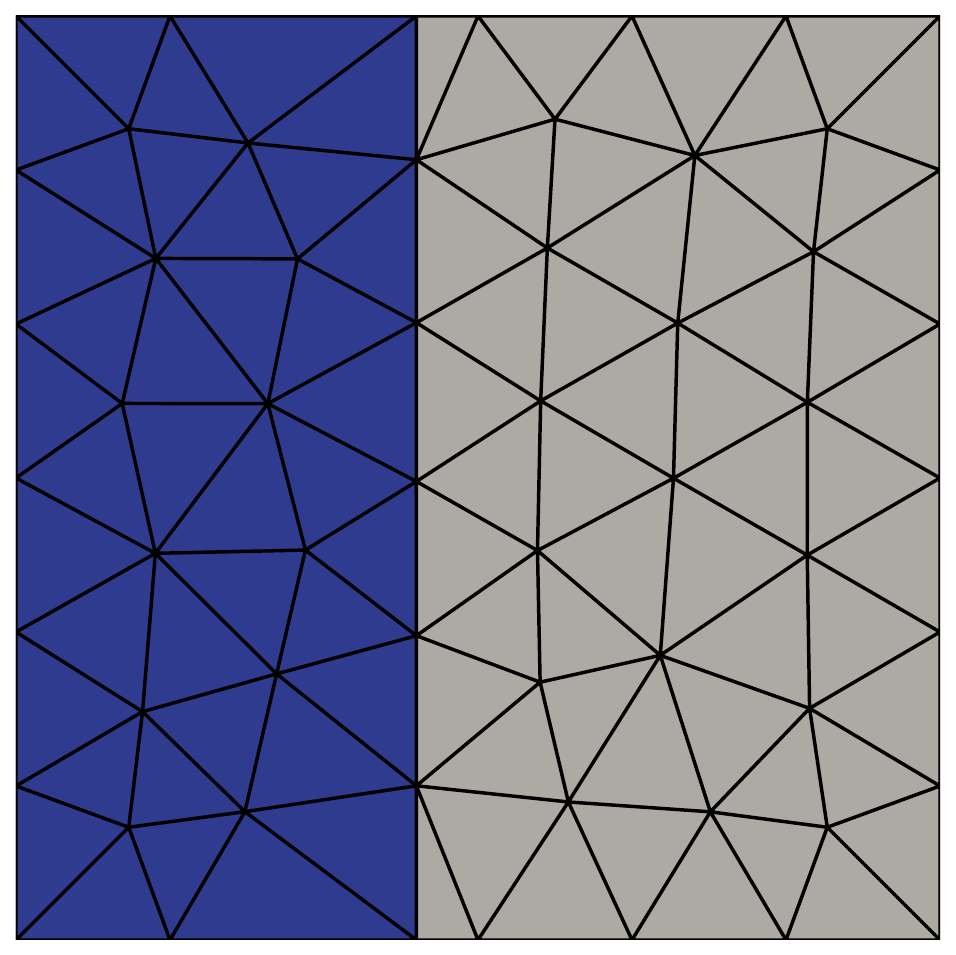} 
    \end{subfigure}
    \hfill
    \begin{subfigure}[b]{0.32\textwidth}  
        \centering 
        \includegraphics[width=\textwidth]{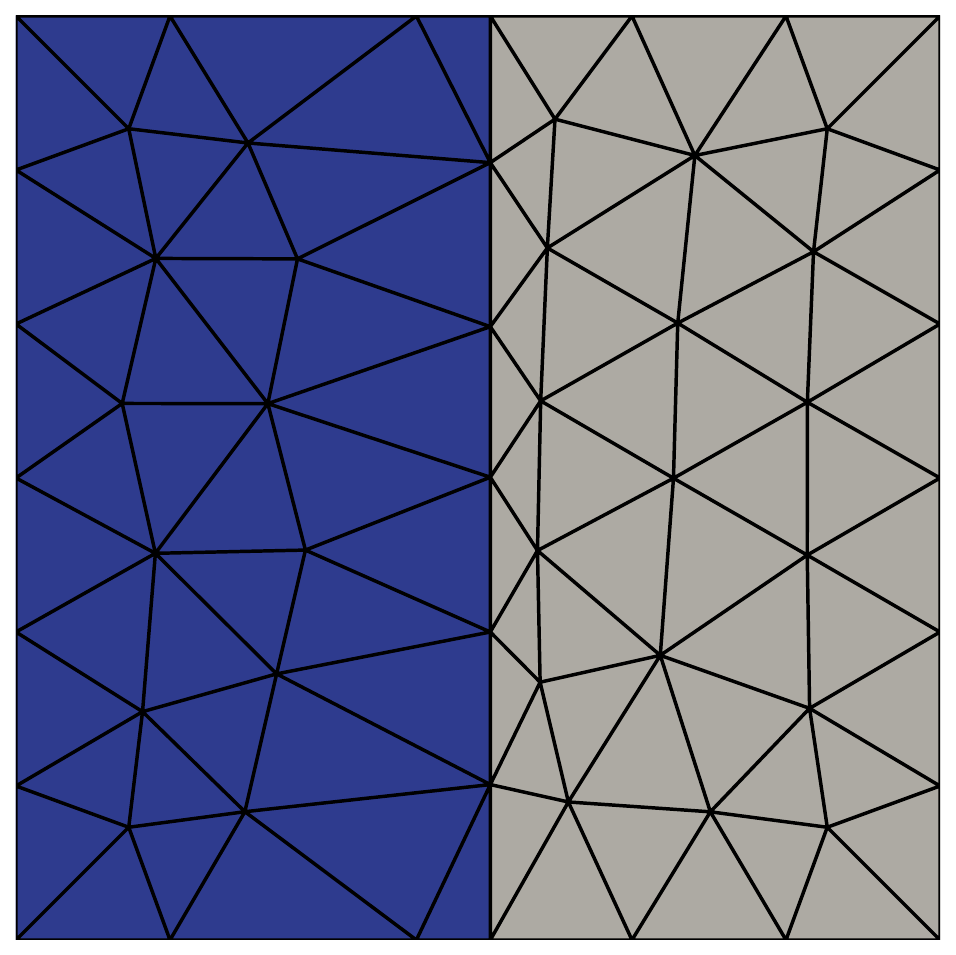} 
    \end{subfigure}
    \hfill
    \begin{subfigure}[b]{0.32\textwidth}   
        \centering 
        \includegraphics[width=\textwidth]{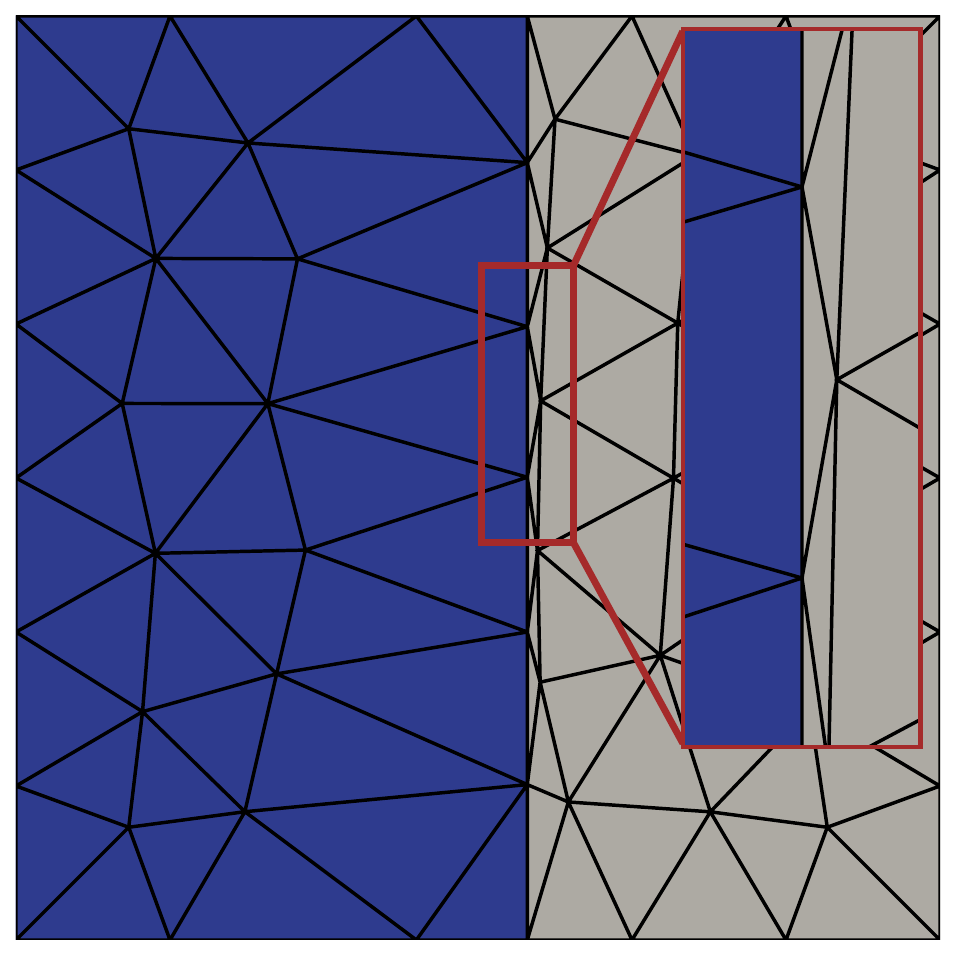}  
    \end{subfigure}
    \\
    \begin{subfigure}[b]{0.32\textwidth}
        \centering
        \includegraphics[width=\textwidth]{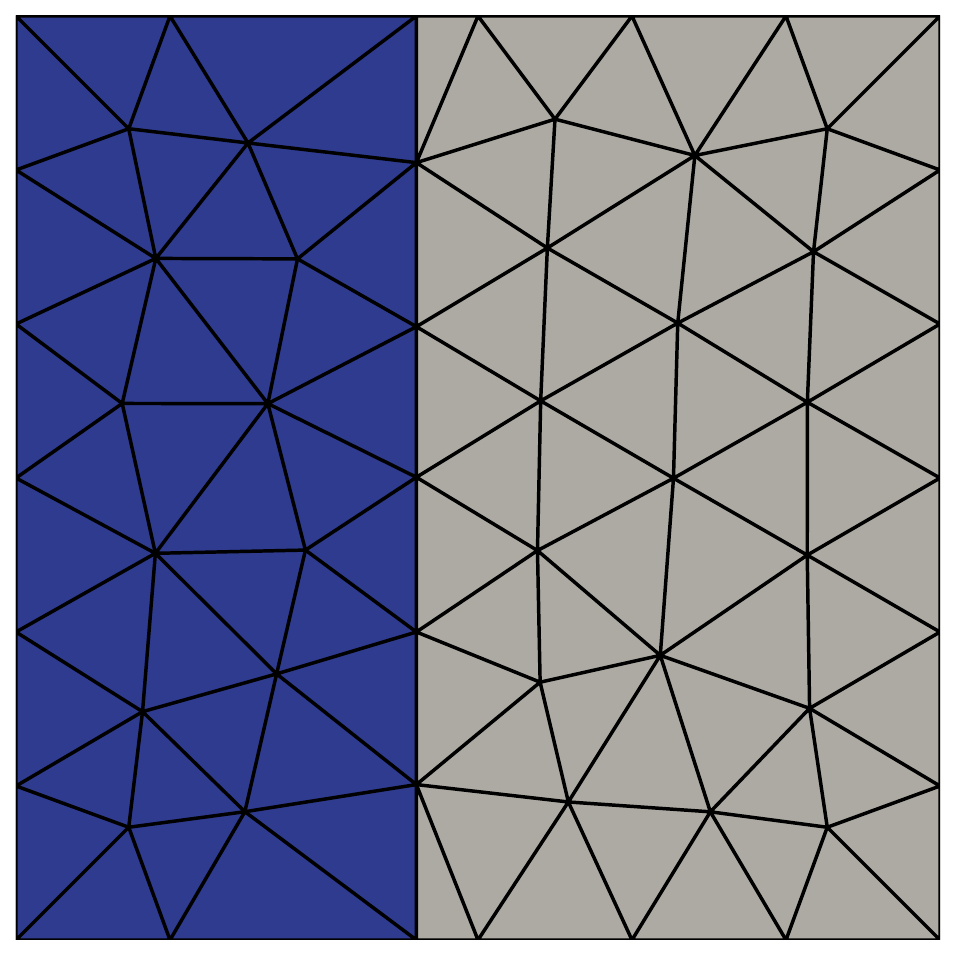}  
    \end{subfigure}
    \hfill
    \begin{subfigure}[b]{0.32\textwidth}  
        \centering 
        \includegraphics[width=\textwidth]{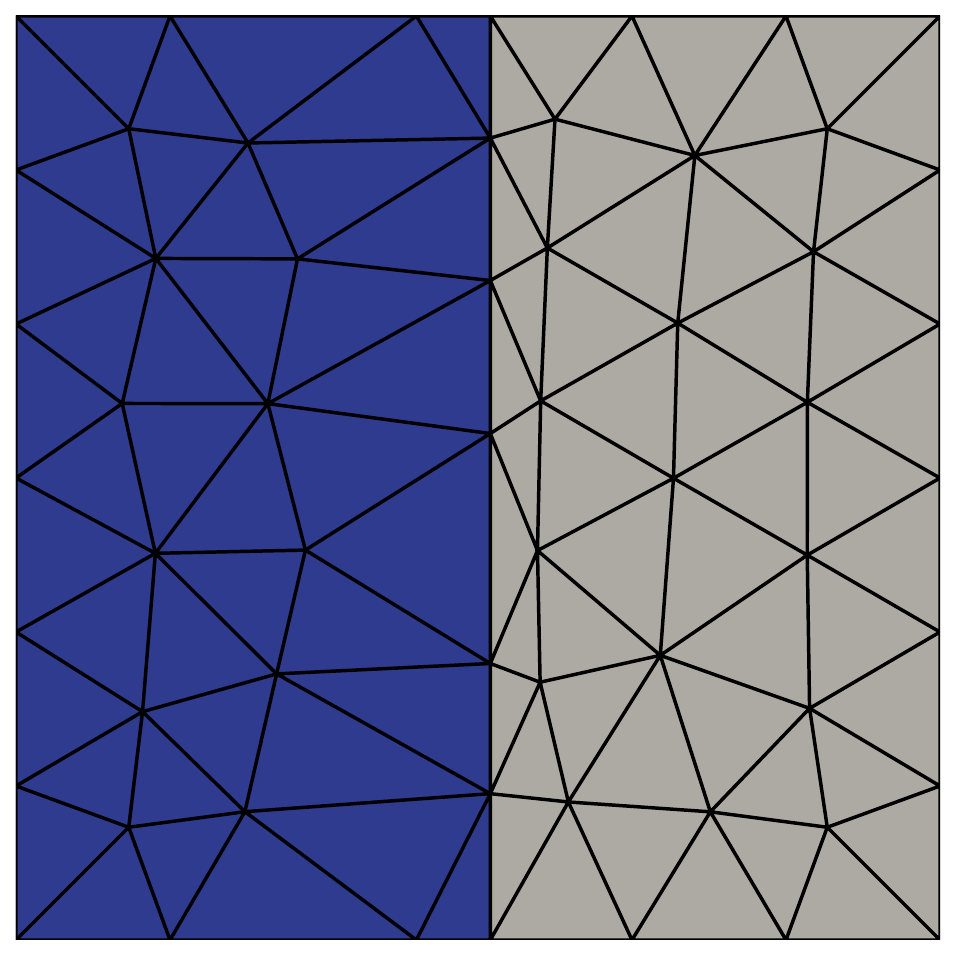}  
    \end{subfigure}
    \hfill
    \begin{subfigure}[b]{0.32\textwidth}   
        \centering 
        \includegraphics[width=\textwidth]{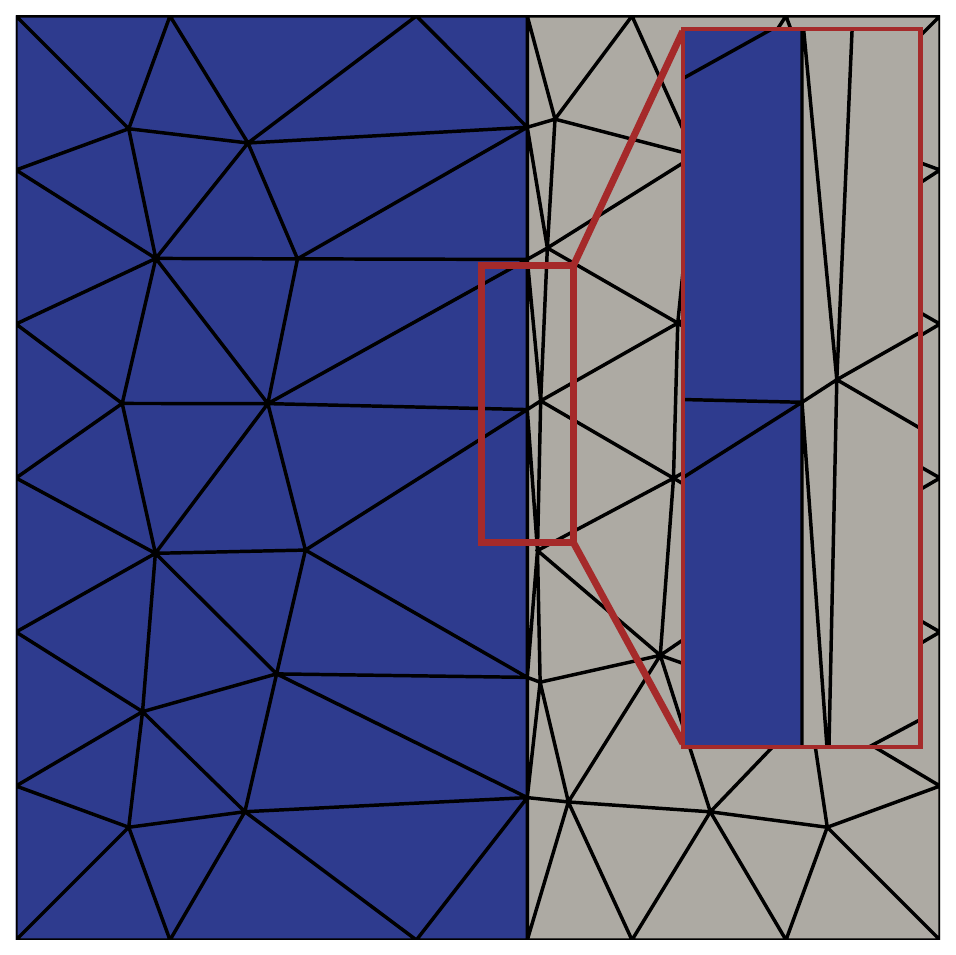}   
    \end{subfigure}
    \caption{Moving the nodes along their velocity (top) or along the edges (bottom).} 
    \label{fig:algo}
\end{figure}

Our mesh deformation/relaying algorithm has from the beginning taken into account this constraint of not creating bands of caps.
Assume a triangular mesh that exactly represent a front at a given time $t$ (Top-left image of Figure \ref{fig:algo}). The front moves to the
right of a constant speed $v$ during a time $\Delta t$ and thus we can move front vertices to the right of $v \Delta t$
(Top-center image of Figure \ref{fig:algo}) and then of $v \Delta t$ again (Top-right image of Figure \ref{fig:algo}). At that point, one
sees that the relay will essentially occur whith triangles that have the form of a cap and that caps are arranged in bands. Such an algorithm
clearly leads to stability problems and should be avoided.
In general, if we are not careful, a mesh deformation algorithm will naturally tend to create these bands of caps.

What if now mesh vertices move along edges of the current mesh (Bottom images of Figure \ref{fig:algo}). Then, a majority of needles will be
created. Isolated caps may be created but their number only depends on the difference between the number of vertices upstream and downstream
the front. Another ingredient of the algorithm is to move the vertices only upstream of the front. This allows the relay process to run smoothly, without having downstream vertices moving in the opposite direction of the front movement and creating some kind of \emph{traffic jam} \cite{X-Mesh}.
Finally, when a vertex of the front has passed the relay to another vertex, it progressively returns to its initial position which allows to maintain a quality mesh upstream of the front.

%A very simple algorithm would be to move the nodes according to their speeds in the manner of front-tracking methods and relay the interface between the nodes to allow for topology changes and large interface moves without remeshing. 

%The first condition found for covergence was the so called minimum angle condition which stated that one should avoid elements whose angle tends to 0 (needles and caps). This condition is only sufficient and has been relaxed in the course of research. First by Babu{\v{s}}ka and Aziz \cite{babuvska1976angle} who weakened the constraint to elements which tend to an angle of $\pi$ radians (the caps). \\

\subsubsection{Front relaying}

\begin{figure}[!t]
    \captionsetup{font=normal}
    \centering
    \begin{subfigure}[b]{0.32\textwidth}
        \centering
        \includegraphics[width=\textwidth]{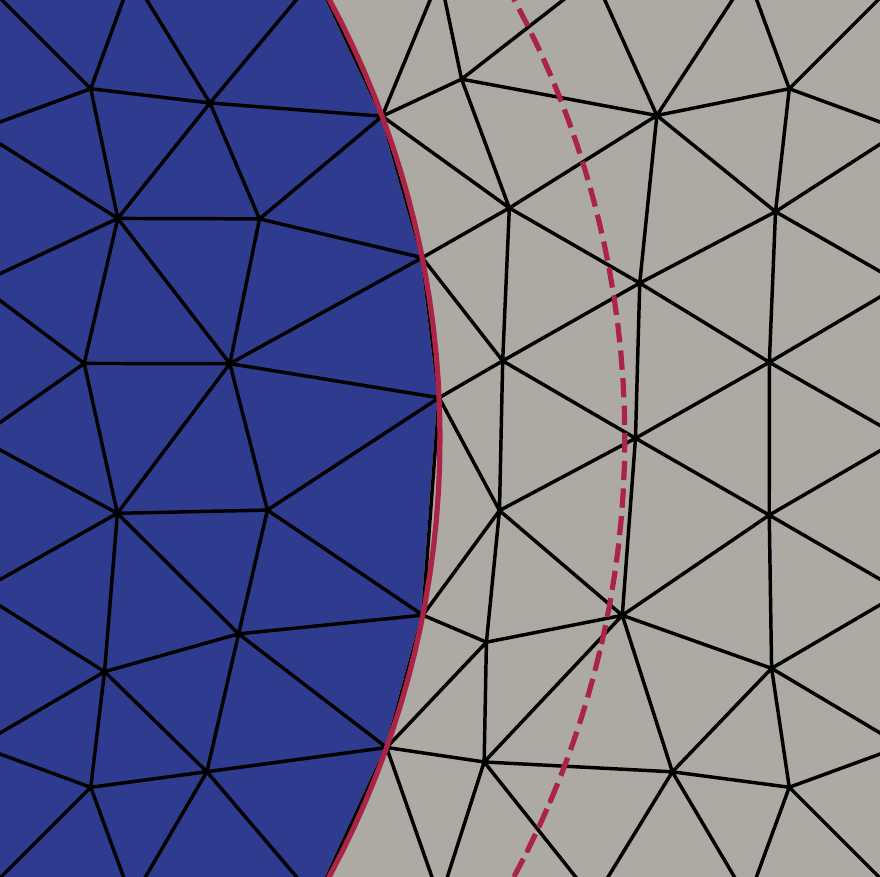}
        \subcaption{Phases at time $t$}  
    \end{subfigure}
    \hfill
    \begin{subfigure}[b]{0.32\textwidth}  
        \centering 
        \includegraphics[width=\textwidth]{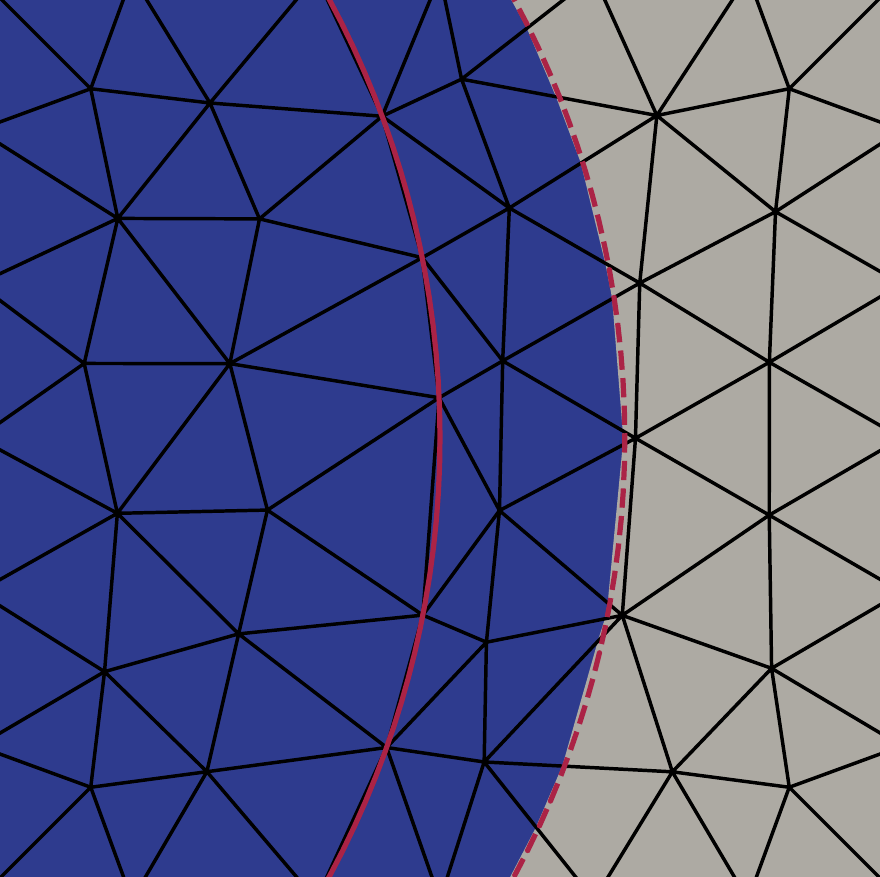}
        \subcaption{Phases at time $t+\Delta t$}    
    \end{subfigure}
    \hfill
    \begin{subfigure}[b]{0.32\textwidth}   
        \centering 
        \includegraphics[width=\textwidth]{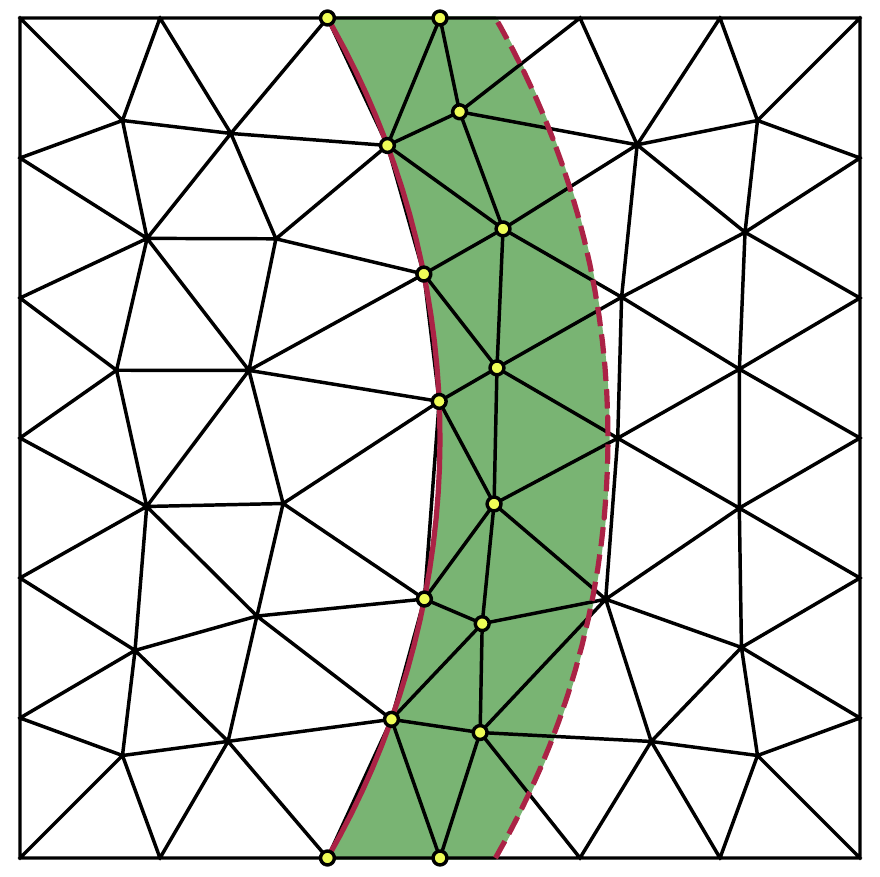}
        \subcaption{Active vertices}    
    \end{subfigure}
    \\
    \begin{subfigure}[b]{0.32\textwidth}
        \centering
        \includegraphics[width=\textwidth]{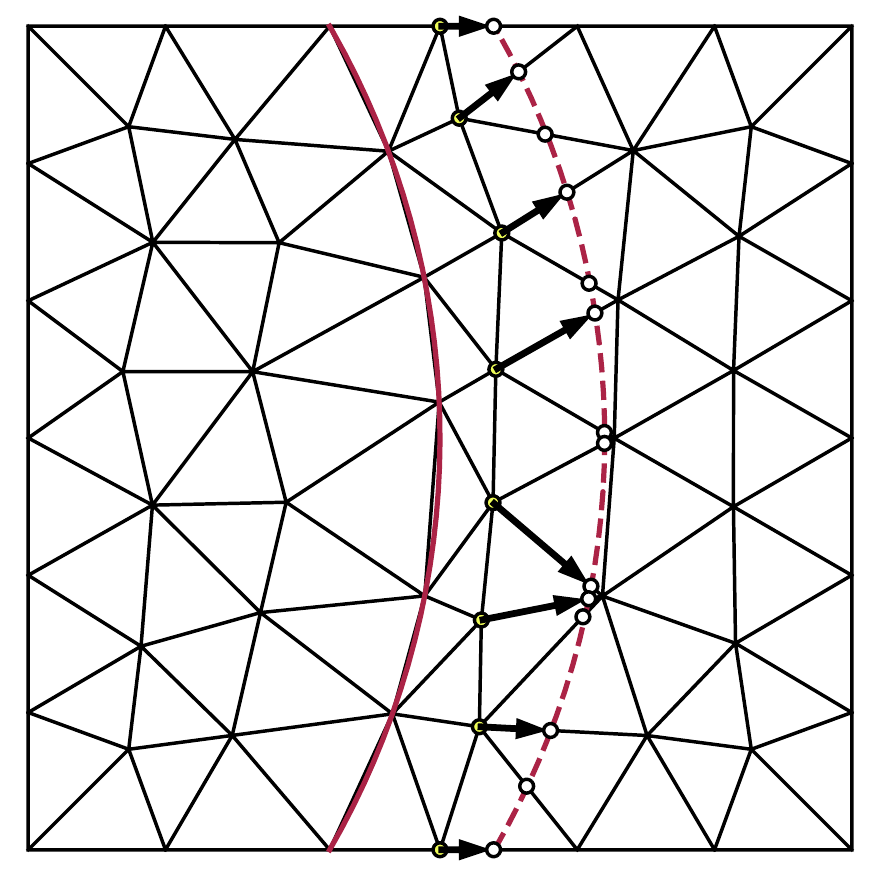}
        \subcaption{Potential targets and moving vertices}    
    \end{subfigure}
    \hspace{0.15\textwidth}
    \begin{subfigure}[b]{0.32\textwidth}  
        \centering 
        \includegraphics[width=\textwidth]{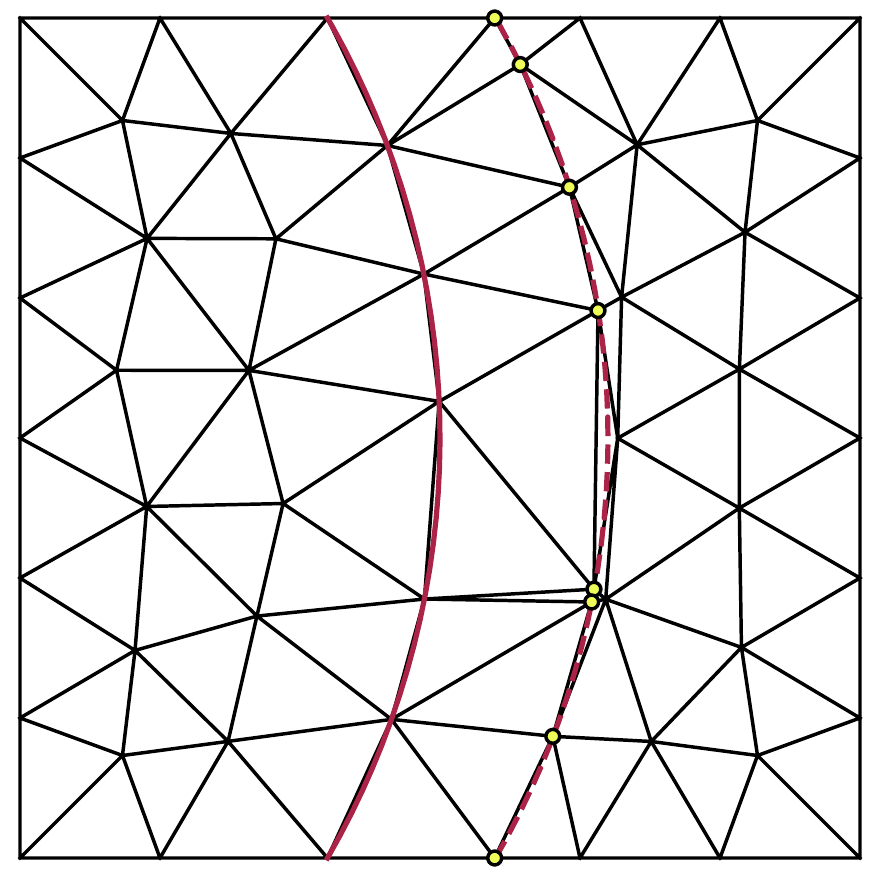}
        \subcaption{New mesh and new front vertices}    
    \end{subfigure}
    \hfill
    \caption{Mesh deformation algorithm.} 
    \label{fig:xmesh6}
\end{figure}

Figure \ref{fig:xmesh6} shows the algorithm used to deform the mesh between 2 time steps so that it continuously conforms to the new interface.
This interface is defined by $\phi=0$ and the sign of $\phi$ determines the phase of the fluid.
On images $(a)$ and $(b)$ of Figure \ref{fig:xmesh6}, the phases of the fluids for 2 consecutive time steps $t$ and $t + \Delta t$ are represented in blue and grey.
The idea of X-Mesh is that the mesh at the two times $t$ and $t+\Delta t$ must conform to the interface.
Image $(b)$ shows the phases at time $t+\Delta t$ on the mesh at time $t$. The new interface at time $t+\Delta t$ -- in dashed lines -- is not conforming
to the mesh. 

% The interface at the first time step is represented in solid line while the interface at the next time step is in dashed line.

The objective of the front relaying algorithm is to deform the existing mesh at time $t$ so that the mesh is conformal to the new interface: we want the interface at time $t+\Delta t$ to be
entirely represented by the edges of the mesh. Additionally, the algorithm must limit as much as possible the creation of bands of caps for stability reasons.
As mentioned above, moving the vertices along the edges of the mesh allows to avoid the creation of bands of caps.
We also want to avoid vertices to accumulate downstream the front. Thus, only nodes
upstream of the front should move towards the front. This starting point allows us to develop a robust front relaying algorithm.
The region in green of image $(c)$ of  Figure  \ref{fig:xmesh6} represent the region where phases have changed between $t$ and $t+\Delta t$.
This region is thus upstream of the front and vertices that move should be chosen in this region. We call \emph{active vertices} the vertices for which the sign
of the level set $\phi$ has changed during the time step. Active vertices include vertices of the previous front which had a level set value of 0 and are coloured in yellow in image $(c)$ of  Figure  \ref{fig:xmesh6}. 

An edge $(ij)$ is a cut-edge if, at time step $t$, its two nodes have a different sign for the level set $\phi$ at time
$t+\Delta t$. There is therefore a potential target on the cut-edge $(ij)$ where the level set is null, i.e. where the new front
will be located at $t+\Delta t$. All active vertices that belong to cut-edges will move: we call them moving vertices
(see image $(d)$ of  Figure  \ref{fig:xmesh6}). Now a moving vertex can belong to more than one cut-edges so we have to
choose one of the potential targets. The moving vertices at the boundary of the domain (the highest and lowest in Figure (4d)) has to be moved on a potential target located on the boundary to preserve the shape of the computational domain. Note that only polygonal computational domains are considered in this work, non-trivial geometries would require special treatment. For the moving vertices inside the domain, we experimentally observed better mesh quality when we chose the potential target that is the closest to the node (arrows in Figure (4d)).
Image $(e)$ of  Figure  \ref{fig:xmesh6}) shows mesh at time $t+\Delta t$ that is conforming to the new interface.

This mesh deformation method is presented in Algorithm \ref{algo_move}. In this algorithm, \textit{levelset} and \textit{previous\_levelset} correspond to the new and old values of the level set. A table called \textit{front} is used to store a boolean for each node to indicate whether it was on the previous interface or not. A second table, \textit{on\_bnd}, is used to specify if the node is on the boundary of the domain and then requires extra attention.\\
If the node is not on the boundary, the function \textit{move\_bulk} is called, in which we find all the neighbours of node i through the function \textit{get\_neighbours}. We can then compute the position of the closest target and the associated neighbour with the function \textit{determine\_closest\_target(i, neighbours)}. These targets corresponds to the position of the 0-value levelset on the edges of the mesh between the node i and the neighbours node.\\
In the case of the boundary nodes, we call \textit{move\_boundary} and we have to distinguish between the neighbours on the boundary and the neighbours on the mass. If possible, we place the node i on a target positioned on the boundary to preserve the domain boundaries. If this is not possible, we look at the neighbours inside the bulk and calculate the closest target. The corresponding neighbour is then placed on that target.
\begin{algorithm}[H] \label{algo_move}
\SetKwInput{KwInput}{Input}                % Set the Input
\SetKwInput{KwOutput}{Output}              % set the Output
\DontPrintSemicolon
  \vspace{0.2cm}

% Set Function Names
  \SetKwFunction{Fbnd}{move\_boundary}
  \SetKwFunction{Fmove}{move\_front}
  \SetKwFunction{Fbulk}{move\_bulk}
 \tcc{{\small The function determine\_closest\_target(i,neighbours) returns the position of the closest levelset 0-value target on the edges between node i and the neighbours.}}
% Write Function with word ``Function''
  \SetKwProg{Fn}{Function}{:}{}
  \Fn{\Fmove}{
        \ForEach{\upshape node i at $\boldsymbol{x}_i$}{
            \If{\upshape front[i] \textbf{is} True \textbf{and} sign(previous\_levelset[i]) $\neq$ sign(levelset[i])}{
                active[i] $\gets$ True
            }
        }

        \ForEach{\upshape node i at $\boldsymbol{x}_i$}{
            \If{\upshape active[i] \textbf{is} True}{
                \If{\upshape on\_bnd[i] \textbf{is} True}{
                    \Fbnd(i)
                } \Else{
                    \Fbulk(i)
                }
            }
        }
       
  }
    \vspace{0.4cm}
  \Fn{\Fbulk \upshape(i)}{
    neighbours $\gets$ get\_neighbours(i)\\
    $\boldsymbol{x}_{\text{closest}}$, neighbour $\gets$ determine\_closest\_target(i,neighbours) \\
    \If{\upshape $\boldsymbol{x}_{\text{closest}}$ \textbf{is not} Null}{
    $\boldsymbol{x}^{\text{new}}_{\text{i}}$ $\gets$ $\boldsymbol{x}_{\text{best}}$\\
    front$^{\text{new}}$[i] $\gets$ True
    }
  }
    \vspace{0.4cm}
  \Fn{\Fbnd}{
    bnd\_neighbours $\gets$ get\_bnd\_neighbours(i)\\
    bulk\_neighbours $\gets$ get\_bulk\_neighbours(i)\\
    $\boldsymbol{x}_{\text{closest}}$, neighbour $\gets$ determine\_closest\_target(i,bnd\_neighbours) \\
    \If{\upshape $\boldsymbol{x}_{\text{closest}}$ \textbf{is not} Null}{
        $\boldsymbol{x}_{\text{i}}$ $\gets$ $\boldsymbol{x}_{\text{closest}}$\\
        front$^{\text{new}}$[i] $\gets$ True\\
    } \Else{
        $\boldsymbol{x}_{\text{closest}}$, neighbour $\gets$ determine\_closest\_target(i,bulk\_neighbours)\\
        \If{\upshape $\boldsymbol{x}_{\text{closest}}$ \textbf{is not} Null}{
            $\boldsymbol{x}_{\text{neighbour}}$ $\gets$ $\boldsymbol{x}_{\text{closest}}$\\
            front$^{\text{new}}$[neighbour] $\gets$ True\\
        }
    }
    
  }

  \caption{Move front algorithm}
\end{algorithm}

A Python code implementing this algorithm is provided as supplementary material to this paper. A small test case of a circular boundary moving along the path of a spiral and coming back allows to see how the algorithm performs for classical cases inside the domain, but also when it reaches the boundaries of the mesh.

A last ingredient of the front relaying algorithm is to progressively move vertices that have left the front to their original position. These nodes are moved by simply following the equation below:
\begin{equation*}
    \mathbf{x}_{n+1} = (1 - \alpha) \mathbf{x}_{n} + \alpha \mathbf{x}_0
\end{equation*}
Where $\mathbf{x}_0$ is the position of the node in the initial mesh (before any mesh deformation) and $\alpha$ is a user parameter between 0 and 1 that changes the speed at which the nodes return to their initial position.
This feature allows to restore a nice mesh upstream the front when the front has passed.\\

The phase of each element can then be determined using the level set function. As the mesh is conformal to the boundary, the iso-value zero of the level set is represented by the edges between the front nodes. Thus, no mesh edge can cut the front after mesh deformation. In this way, no elements have vertices with a level set value of opposite sign. The phase of each element can thus be obtained from the level set value of its vertices. If the 3 vertices of a triangle are located on the front (and thus have a level set value of 0), the phase of the element is ambiguous. We call these elements triple-zero elements. This occurs when the curvature of the front is important for the characteristic length of the element or when a topological change occurs. A simple way to deal with such elements is to look at their phase at the previous time step. If the element was already a triple-zero element, we keep the same phase as before, so that these elements don't oscillate from one time step to another. If the element was not a triple-zero before, we change the phase of the element. In this way, the merging of the phases can take place.

\subsection{Coupling}
This section describes the algorithm used to couple the different parts presented previously: the resolution of the Navier-Stokes equations, the resolution of the advection equations, the displacement of the mesh with the algorithm proposed for X-Mesh and the redistancing of the level set. All the results of this paper have been obtained with a sequential algorithm presented in Algorithm 1. The simulation of a time step consists in the successive resolution of these different steps. The Navier-Stokes equations are solved to obtain the new velocity field. This velocity field allows us to advect the level set in order to obtain the new level set values on the mesh defined by the positions $\mathbf{x}_n$. The mesh is then deformed to match the new interface defined by the level set. Since the mesh has moved, the level set values at the nodes are no longer correct. We then obtain the new level set for the new positions of the mesh $\mathbf{x}_{n+1}$ by applying the fast-marching method \cite{sethian1996fast} with the nodes of the front in ${n+1}$ as seeds. This coupling has shown a very good robustness in practice for test cases with many topological changes. To increase the precision of the scheme, one could be tempted to iterate between the mesh deformation and the solution of the Navier-Stokes equations. This would allow the mesh deformation between $t_n$ and $t_{n+1}$ to be taken into account directly in the equations between $t_n$ and $t_{n+1}$. This is however not trivial and a simple fixed point method does not converge quickly. The position of the mesh can oscillate between two states without converging, especially when there is a sudden change of configuration e.g. when a topological change occurs.

\begin{algorithm}
\caption{Sequential coupling for one time-step}\label{alg:stag}

$\mathbf{u}_{mesh}$ = $\left(\mathbf{x}_{n} - \mathbf{x}_{n-1} \right) \frac{1}{\Delta t}$\\
$\mathbf{u}_{n+1}$ = \text{Navier-Stokes} $(\mathbf{u}_n, \phi_n, \mathbf{u}_{mesh})$\\
$\phi_{adv}$ = \text{level set} $(\mathbf{u}_{n+1}, \phi_n)$\\
$\mathbf{x}_{n+1}$, $\text{front}_{n+1}$ = \text{X-Mesh} $(\mathbf{x}_n, \phi_{adv}, \phi_n, \text{front}_n)$\\
$\phi_{n+1}$ = \text{Fastmarching} $(\mathbf{x}_{n+1}, \text{front}_{n+1})$

$t$ = $t+\Delta t$ \\

\end{algorithm}

 \section{Surface Tension}
When considering a two-phase flow, depending on the material properties of the two fluids and the problem considered, it is not always possible to neglect surface tension. This is notably the case for a large number of industrial processes where flows related to the dynamics of bubbles appear. A good representation of the surface tension is essential to correctly simulate these flows. From equation \ref{qdm} we can directely link the surface tension to the conservation of momentum condition at the interface. Indeed, when this condition is taken in the direction normal to the interface, it is expressed by:
\begin{align} \nonumber
    \mathbf{n} \cdot \left[ -pI + \mu \hspace{2pt} \frac{1}{2} \left(\nabla \mathbf{u} + \nabla \mathbf{u}^T \right) \right] \cdot \mathbf{n} &= \mathbf{n} \cdot \sigma \kappa \mathbf{n}\\ \nonumber
    [p] &= -\sigma \kappa + \mathbf{n} \cdot \left[\mu \hspace{2pt} \frac{1}{2} \left(\nabla \mathbf{u} + \nabla \mathbf{u}^T \right) \right] \cdot \mathbf{n} \\
    [p] &= -\sigma \kappa
\end{align}
The term $\mathbf{n} \cdot \left[ \mu \hspace{2pt} \frac{1}{2} \left(\nabla \mathbf{u} + \nabla \mathbf{u}^T \right) \right] \cdot \mathbf{n}$ is exactly 0 because both fluids are considered Newtonian and incompressible. It is therefore possible to obtain the value of the pressure drop at the interface based on the curvature of the interface and the surface tension coefficient.
To implement the effects of surface tension, we will base ourselves on the Ghost Fluid Method \cite{popinet2018numerical} \cite{kang2000boundary} by imposing this pressure jump at the interface. In this method on a fixed mesh, the pressure gradient operator $\nabla p$ is modified to take into account the classical pressure in addition to the pressure jump for the elements intercepted by the interface. One could then think that the method is sharp because the pressure jump is well located at the interface however the interface has a thickness of the size of an element. Indeed, the method is insensitive to a displacement of the interface of $\pm \frac{h}{2}$ with $h$ the element size \cite{popinet2018numerical}. 
Our approach is similar to the ghost fluid method in the sense that we also explicitly add the pressure jump to model the surface tension effects. The difference is that with X-Mesh the nodes of the mesh are positioned at the interface, so we need to add the pressure jump directly at the node.
The modified pressure $\bar{p}$ used in the $\nabla p$ operator of equation \ref{NS_discr} is only different for the boundary nodes which will have as value:
\begin{align*}
    \bar{p}_i = \begin{cases}
       p_i \hspace{0.3cm} &\text{if } \Omega_e \in \Omega_1\\
       p_i + \sigma \kappa_i \hspace{0.3cm} &\text{if } \Omega_e \in \Omega_2
    \end{cases}
\end{align*}
with $p_i$ the pressure at node $i$, $\kappa_i$ the interface curvature at node $i$ and $\Omega_e$ the considered element.\\ 
The representation of the surface tension thus obtained is exactly sharp because contrary to the ghost fluid method on fixed mesh the thickness of the interface is zero. An arbitrarily small displacement of the interface induces a displacement of the mesh which modifies the pressure gradient operator and thus the surface tension.

\begin{figure}[!htb]
    \captionsetup{font=normal}
    \centering
    \begin{subfigure}[b]{0.314\textwidth}
        \centering
        \includegraphics[width=\textwidth]{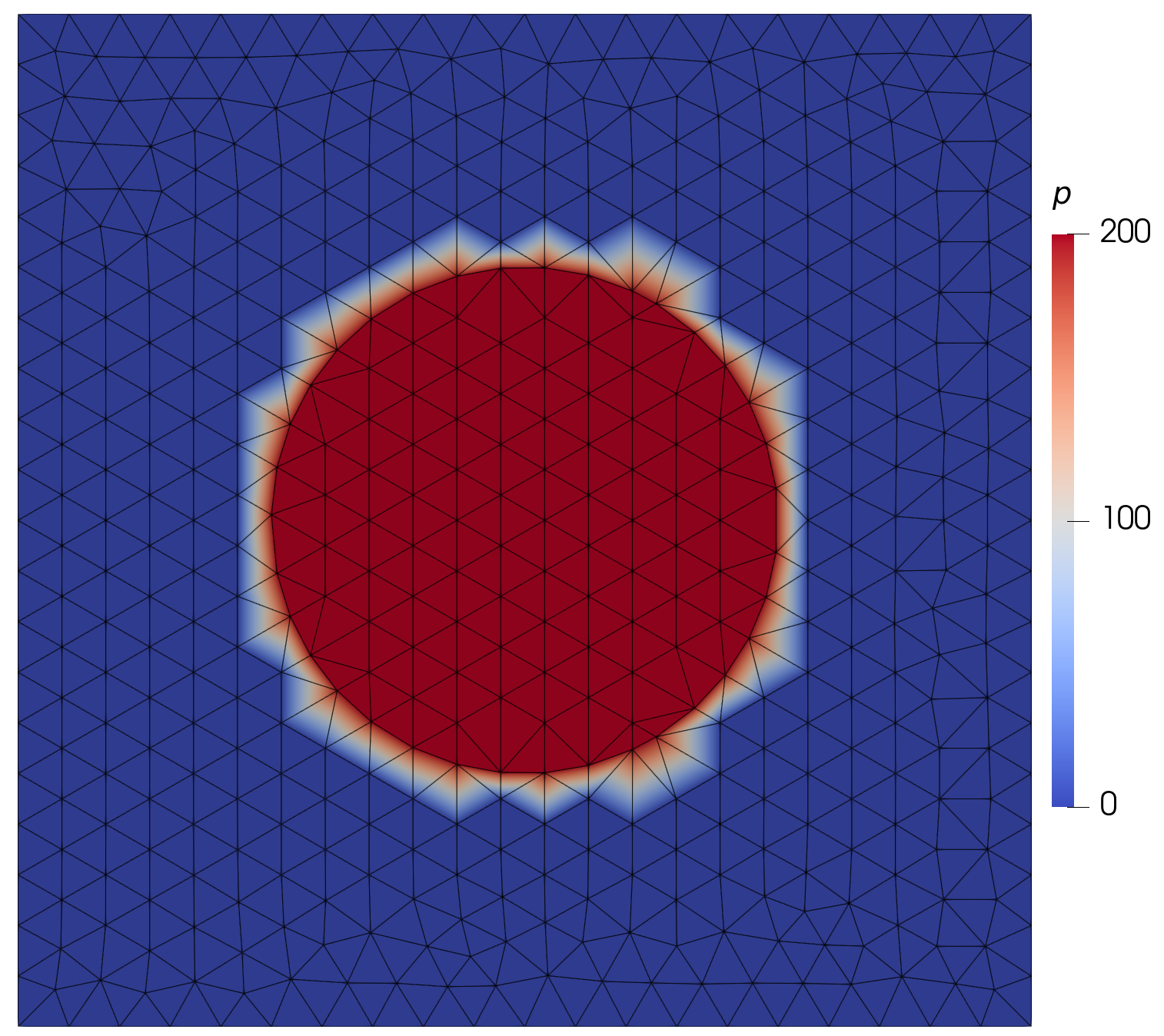}
        \subcaption{Pressure $p$}  
    \end{subfigure}
    \hfill
    \begin{subfigure}[b]{0.314\textwidth}
        \centering
        \includegraphics[width=\textwidth]{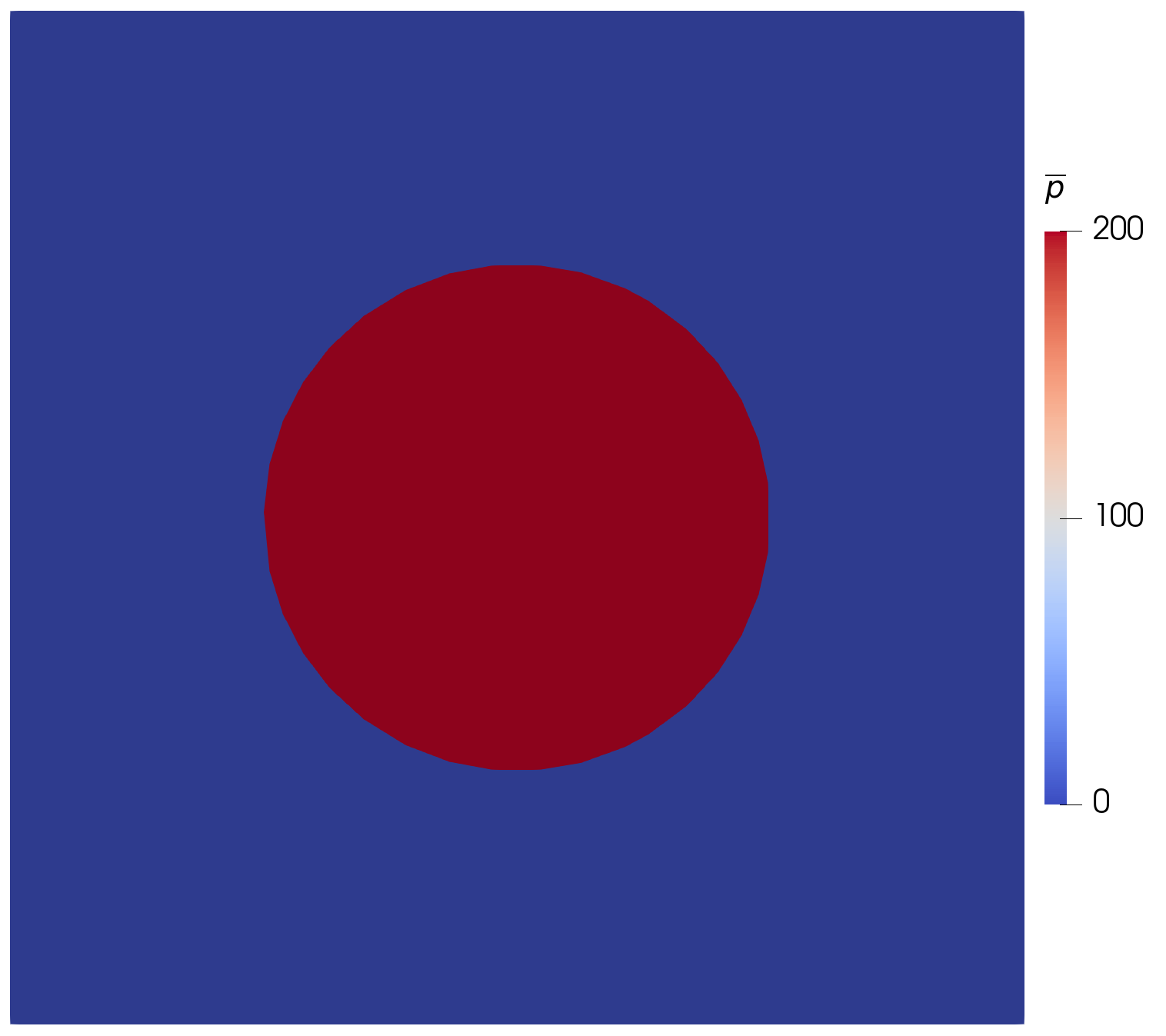}
        \subcaption{Modified pressure $\bar{p}$}  
    \end{subfigure}
    \hfill
    \begin{subfigure}[b]{0.33\textwidth}  
        \centering 
        \includegraphics[width=\textwidth]{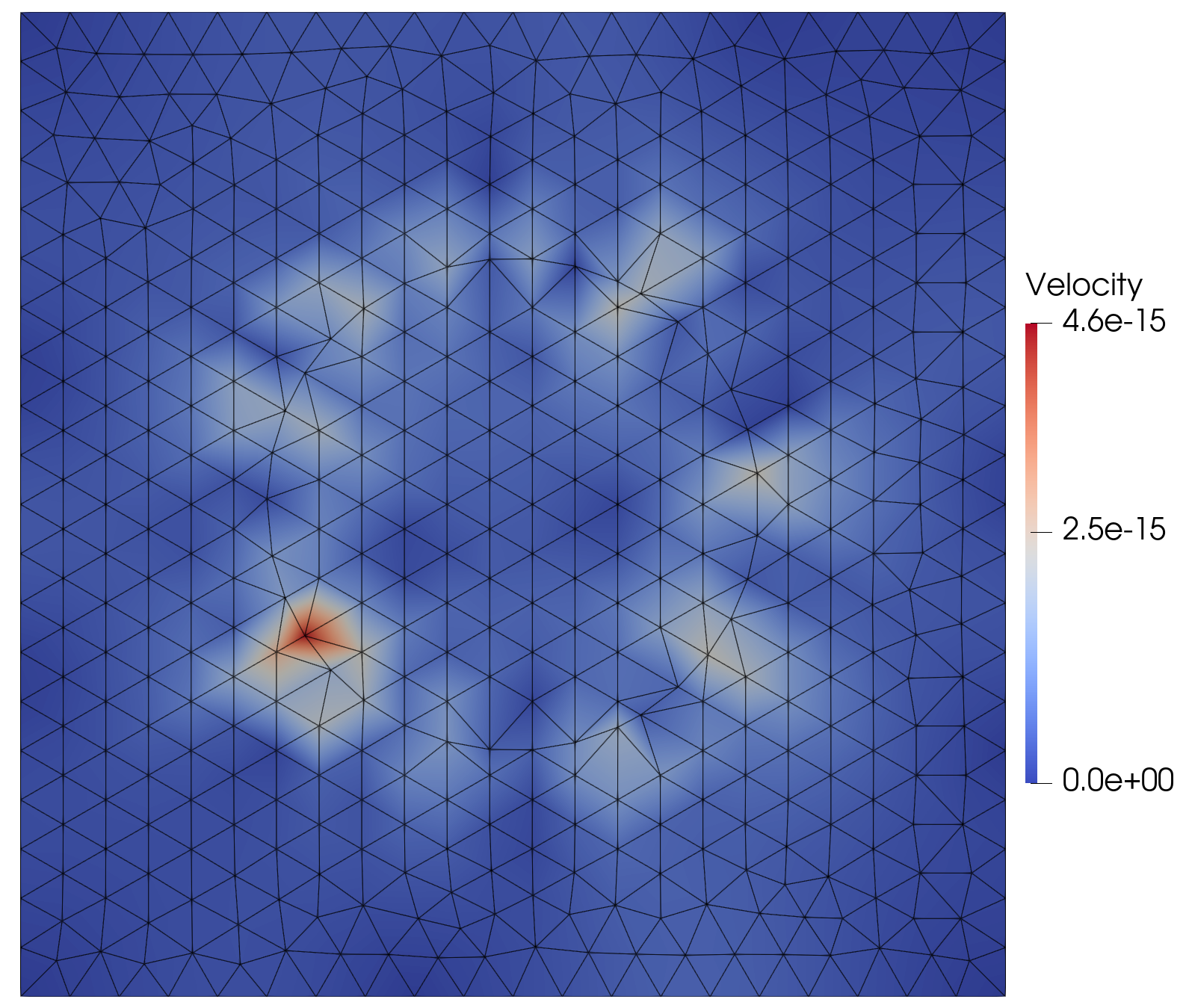}
        \subcaption{Parasitic currents}    
    \end{subfigure}
    \caption{Static bubble.}
    \label{fig:static}
\end{figure}

This pressure jump can be observed in Figure \ref{fig:static} in the case of a static bubble with the classical pressure (a) and the modified pressure (b) which is well constant everywhere except exactly at the interface. When trying to model the surface tension, a very important step is the measurement of the curvature of the interface. As we know the nodes positioned on the interface, it is possible to approximate the curvature by calculating the radius of the circumscribed circle of 3 consecutive nodes belonging to the interface. This very local method has the merit of being exact for the circle, the equilibrium state of the static bubbles.
Figure \ref{fig:static} shows the pressure, the modified pressure and the velocity field for the static Laplace bubble. In this validation, a circular bubble without gravity is simulated. The expected pressure inside the bubble was calculated analytically by Laplace and is worth:
\begin{align*}
    p_{bubble} = \frac{\sigma}{R}
\end{align*}
with $R$ the radius of the bubble. This expression consider that the pressure outside the bubble is zero.\\

This bubble is in equilibrium and the pressure gradient should balance the surface tension effect, but this trivial equilibrium is difficult to reproduce numerically. Due to numerical inaccuracies or poor numerical modelling of the surface tension, spurious or parasitic currents may appear. Figure 5 c is a color map of the velocity norm $|\mathbf{v}|$ obtained with the approach presented in this paper for the static bubble test case. These spurious currents oscillate all around the interface, but the norm of these currents is of the magnitude of $10^{-15}$, the order of machine precision.
The continuous pressure $p$ and the modified (discontinuous) pressure $\bar{p}$ is also represented in the two first figure of \ref{fig:static}. In this case the bubble considered has a radius of size $R = 0.5 \hspace{2pt}[m]$ and the surface tension considered is $\sigma = 100 \hspace{2pt}[N/m]$. The obtained pressure inside the bubble is $p=200 \hspace{2pt}[Pa]$ as expected by the Laplace formula. The time $t_{obs}$ at which the pressure and the velocity field is observed is $250$ times greater than the characteristic time of this problem $t_{char} = \frac{D \mu}{\sigma}$.

\section{Results}
In this section, we test our method by applying it to a validation case of sloshing and to several test cases, namely: viscous dam break, Rayleigh-Taylor instability, single bubble rising and two bubbles merging.
\subsection{Verification of the solver -- sloshing}
The \emph{sloshing problem} consists in computing the free oscillations of a liquid in a tank. This problem essentially allows to verify that our solver solves the right equation because it is one of the few cases where an analytical solution exists in the case of small perturbations of a planar interface \cite{wu2001effect}. In the analytical setting, the interface is sufficiently simple so that it can be represented by a height function $\eta(x,t)$ that is initialized as
\begin{equation*}
    \eta(x, 0) = d + \eta_0 \sin{k \left(0.5 - x \right)}
\end{equation*}
where $d=1.0$ is the mean elevation, $\eta_0 = 0.01$ is the initial oscillation amplitude and $k$ is the wave number.

The analytical evolution of the maximum height of the free surface $\eta$ depend on $\nu = \frac{\mu}{\rho}$ the kinematic viscosity and  the dimensionless parameter $\kappa = \frac{g}{\nu^2 k^3}$  :
\begin{align*}
    \frac{\eta(t)}{\eta_0} = 1 - \kappa e^{-\nu k^2t} f(\nu k^2 t)
\end{align*}
where $f(t)$ is the inverse Laplace transform of the function
\begin{align*}
    F(s) = \frac{1}{(s-1)\left( (s+1)^2 - 4 s^{\frac{1}{2}} + \kappa \right)}.
\end{align*}
The computational domain considered is a rectangle of dimensions $[0,d] \times [0, 1.3d]$ and the four wall have free slip condition.

\begin{figure}
    \centering
    \captionsetup{font=normal}
    \includegraphics[width=0.8\textwidth]{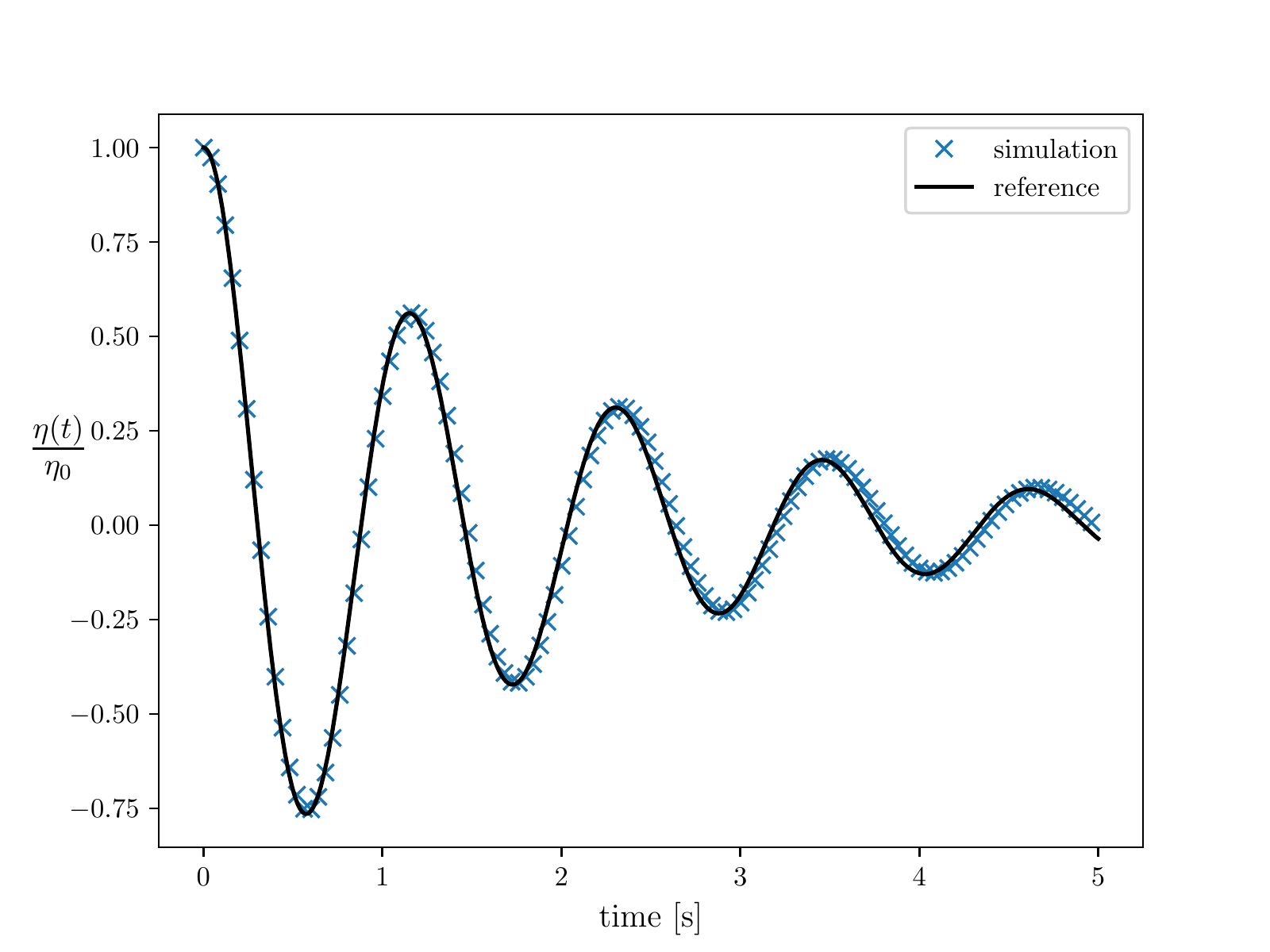}
    \caption{Wave elevation evolution in a sloshing problem.}
    \label{fig:sloshing}
\end{figure}

Figure \ref{fig:sloshing} shows the evolution of the perturbation $\eta$ over time. We can observe that the simulation is in good agreement with the analytical solution. Since this analytical solution is obtained from the linearized Navier-Stokes equations, the initial perturbation $\eta_0=10^{-2}$ cannot be too large with respect to $d=0.5$.

%% FIXME  --> PAS CLAIR !!!!!
%%The mesh used is relatively coarse -- about $3,900$ vertices -- and the relay of the front between the vertices is therefore only done at the beginning of the simulation, the rest is closer to classical ALE method.
 For the reference solution of the linearized Navier-Stokes equations to be valid, the amplitude of the sloshing must be small enough. However, the mesh used for this simulation is quite coarse with $3 900$ vertices. At the beginning of the simulation the characteristic size of the elements is smaller than the amplitude of the sloshing and thus a relay of the front is done between the different vertices. However, quite quickly the amplitude of the sloshing decreases and the simulation is close to a classical ALE simulation without relaying of the interface

\subsection{Dambreak}
The dambreak test case consists of the collapse of a fluid column in a container. It corresponds to the sudden rupture of a dam and the flow of the impounded water. The evolution of the interface in this problem is complex and involves multiple topology changes. Its complexity has made it a reference test case to validate two-phase flow simulations. The considered column has height $H = 0.4 \hspace{2pt}[m]$ and width $L=0.4 \hspace{2pt} [m]$. It collapses in a calculation domain of size $1.4 \times 1.1 \hspace{2pt} [m^2]$ . Boundary conditions of free-slip are applied to the 4 walls. The two fluids used in this test case can be described by their density $\rho$ and kinematic viscosity $\nu$:
\begin{align*}
    \rho_0 = 10 \hspace{2cm} \rho_1 = 1000 \hspace{2cm} \nu_0 = \nu_1 = 10^{-3}
\end{align*}

\begin{figure}[!ht]
    \centering
    \captionsetup{font=normal}
    \begin{subfigure}[b]{0.32\textwidth}
        \centering
        \includegraphics[width=\textwidth]{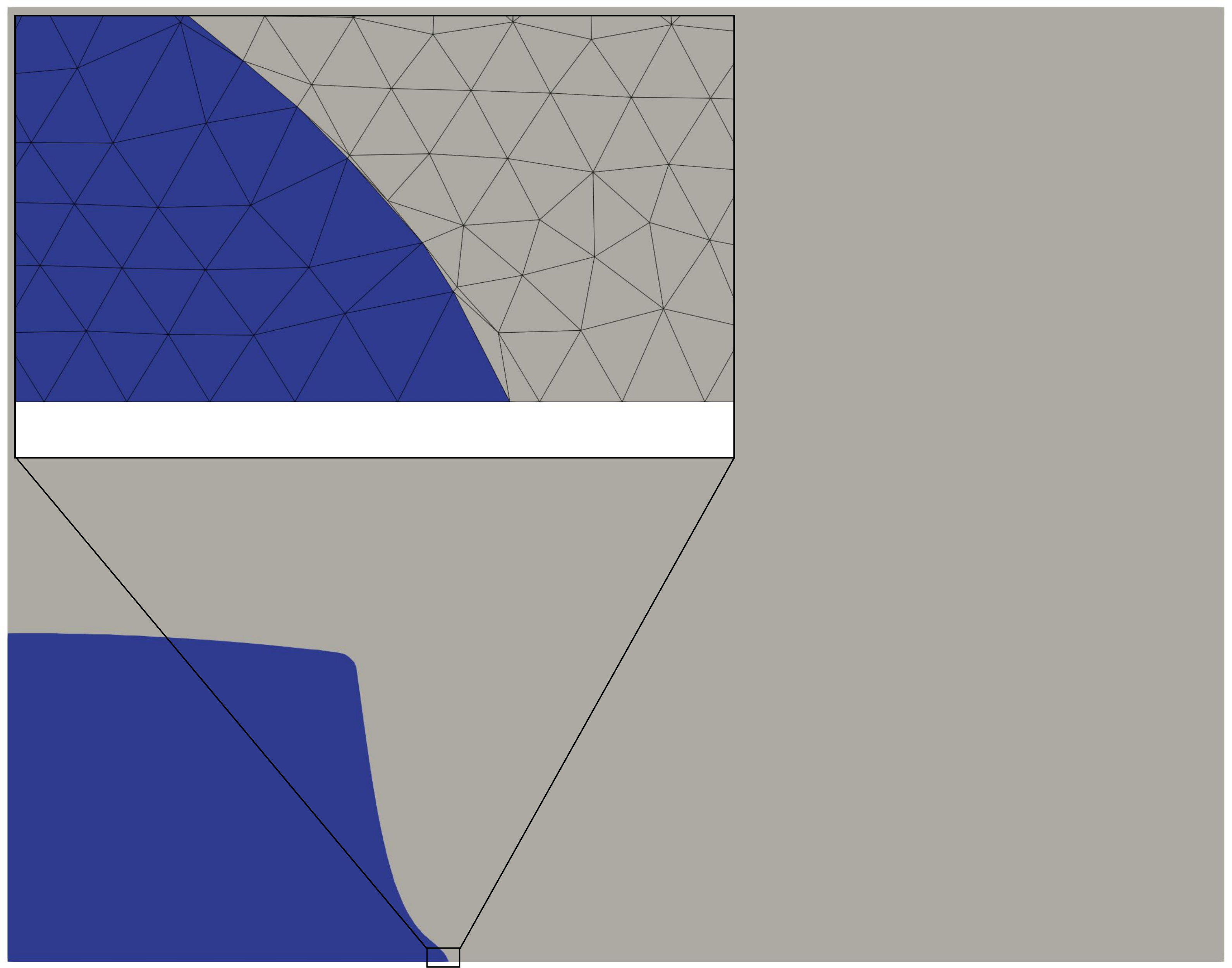}
    \end{subfigure}
    \hfill
    \begin{subfigure}[b]{0.32\textwidth}  
        \centering 
        \includegraphics[width=\textwidth]{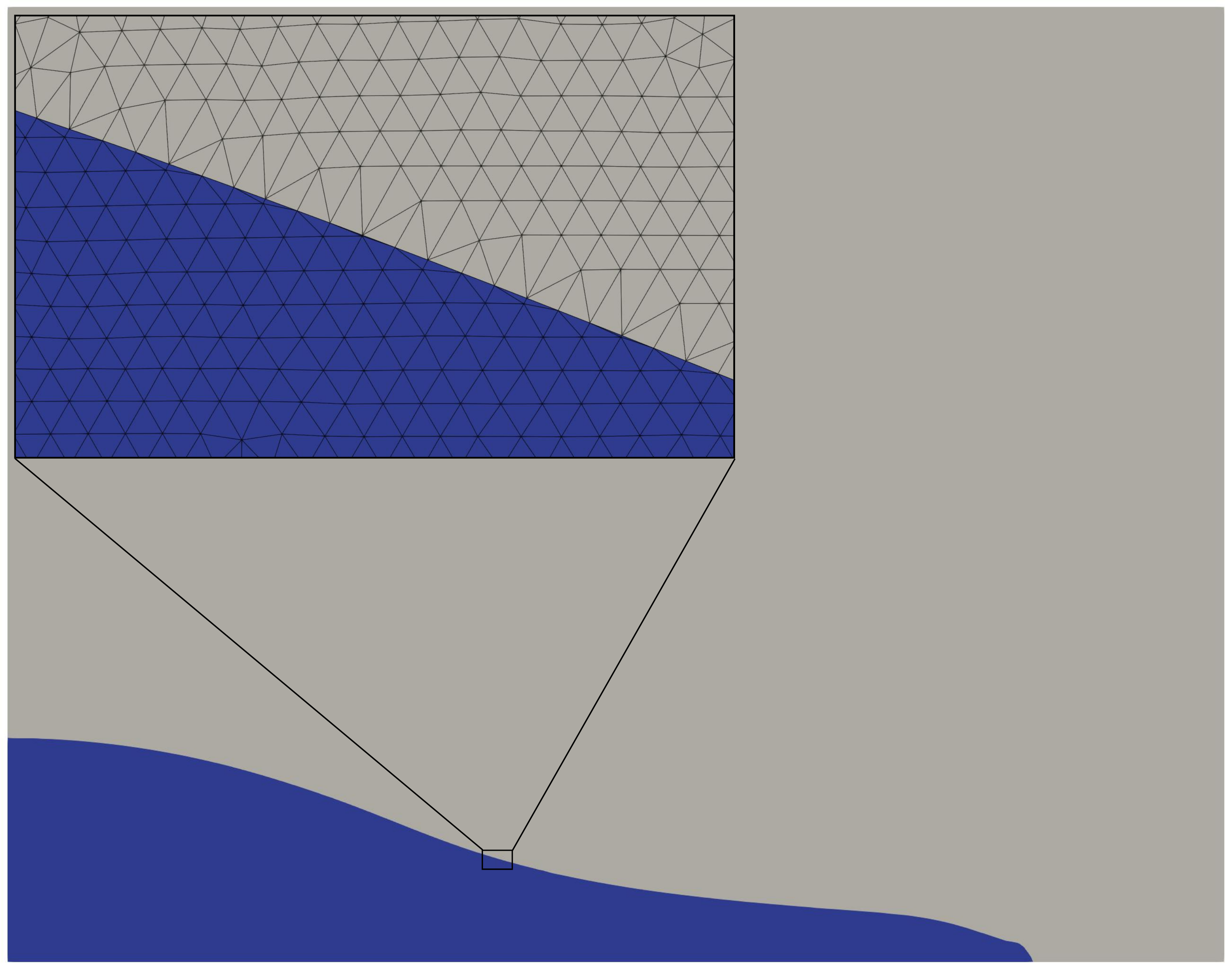}
    \end{subfigure}
    \hfill
    \begin{subfigure}[b]{0.32\textwidth}   
        \centering 
        \includegraphics[width=\textwidth]{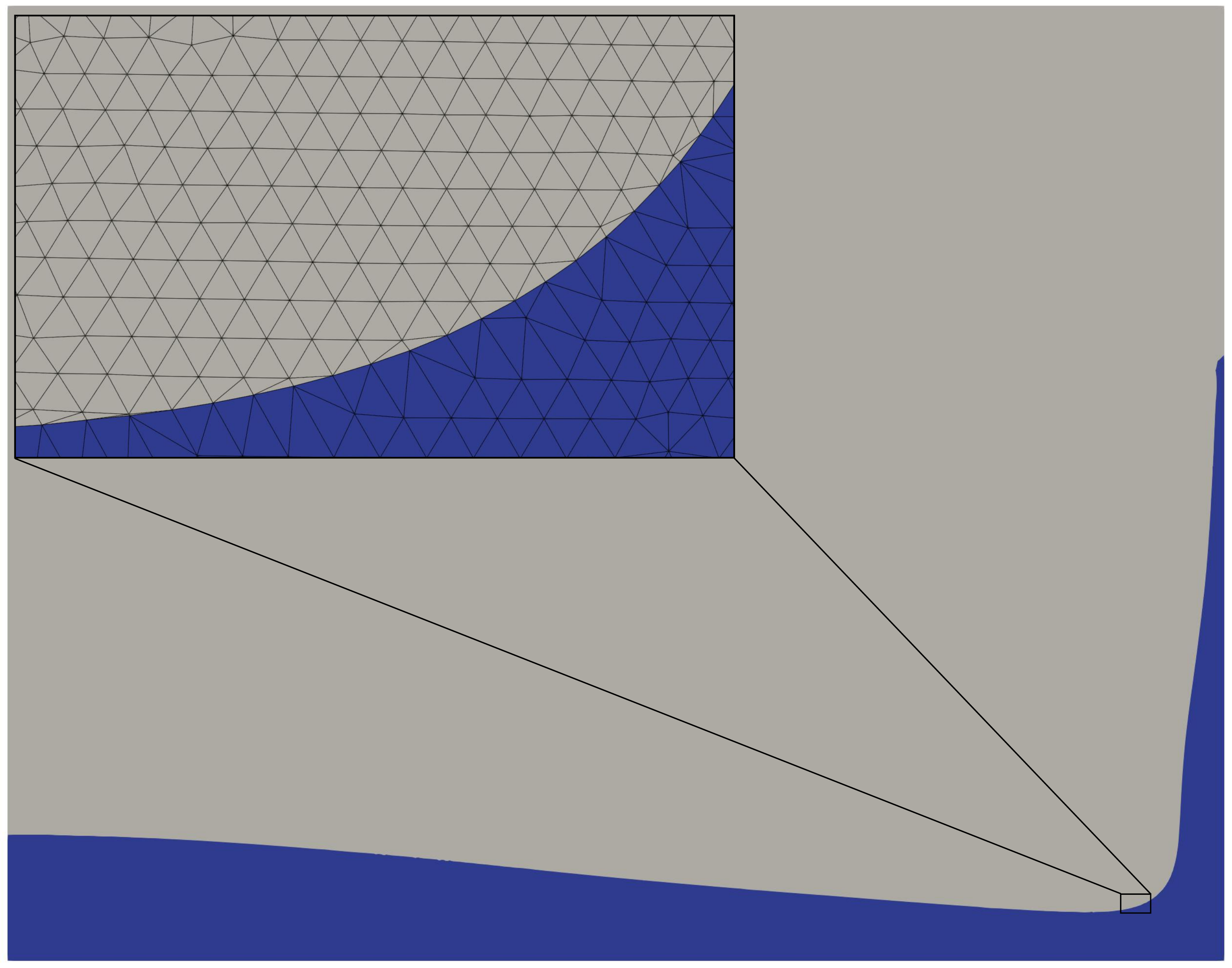}
    \end{subfigure}
    \\
    \begin{subfigure}[b]{0.32\textwidth}
        \centering
        \includegraphics[width=\textwidth]{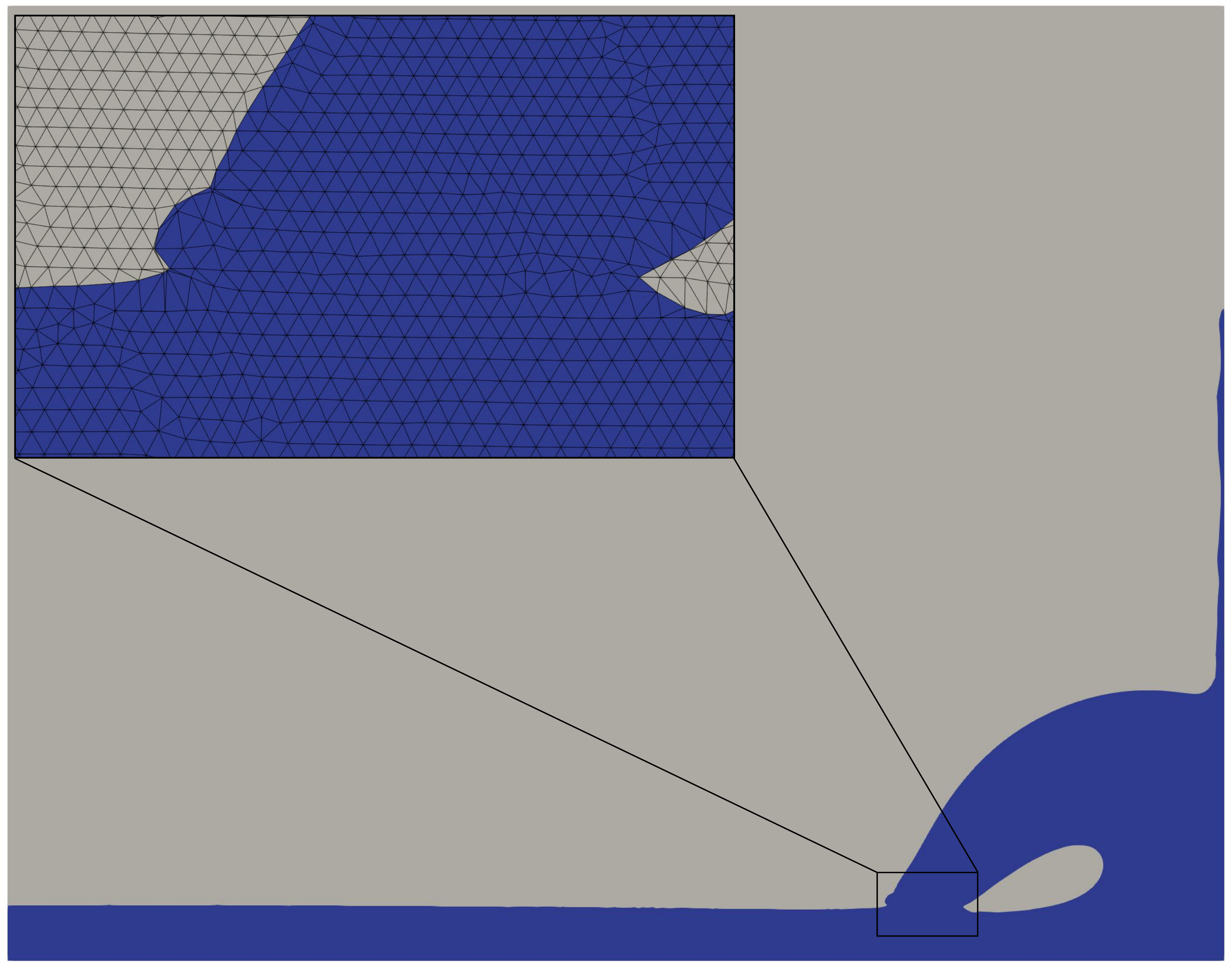}
    \end{subfigure}
    \hfill
    \begin{subfigure}[b]{0.32\textwidth}  
        \centering 
        \includegraphics[width=\textwidth]{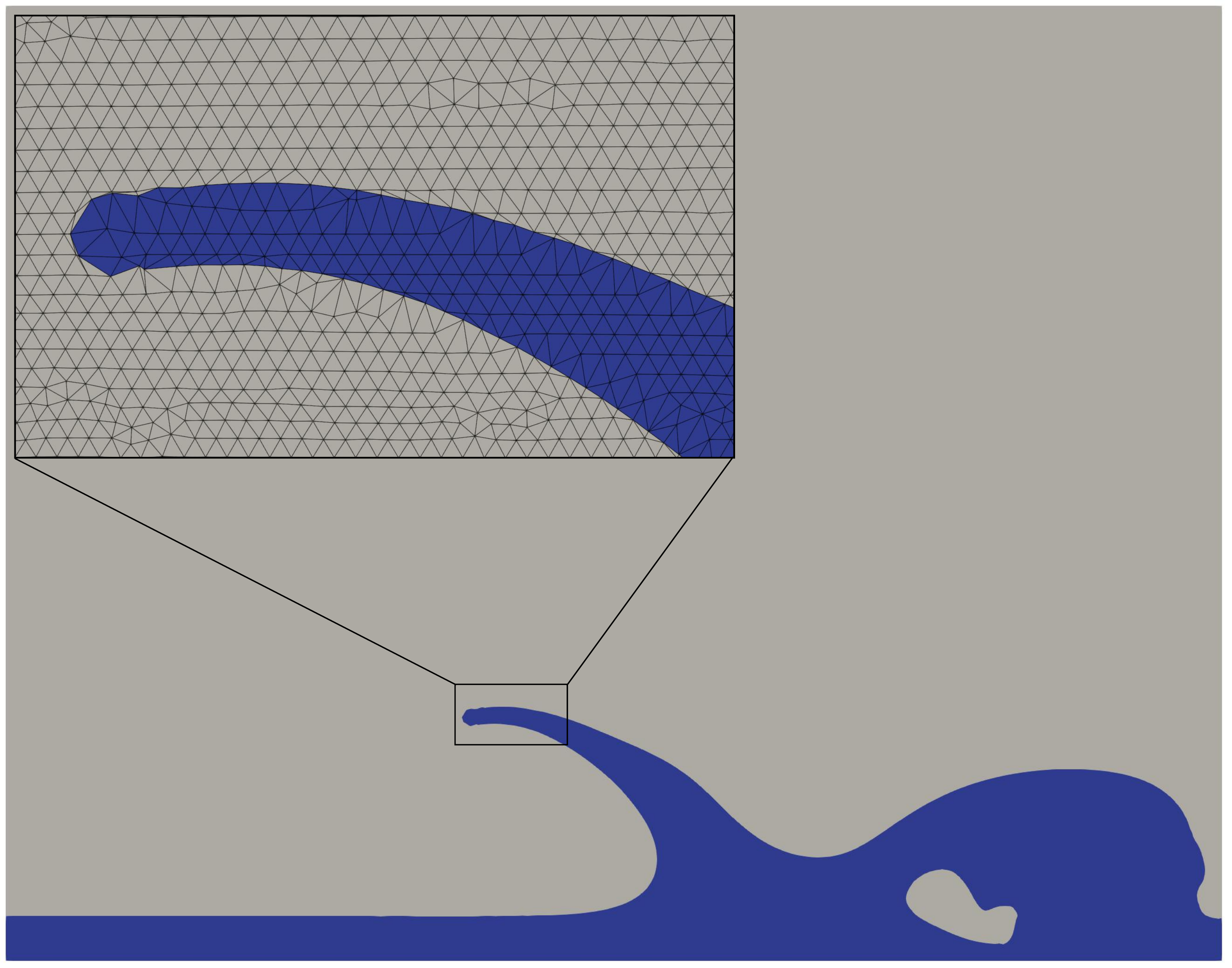}
    \end{subfigure}
    \hfill
    \begin{subfigure}[b]{0.32\textwidth}   
        \centering 
        \includegraphics[width=\textwidth]{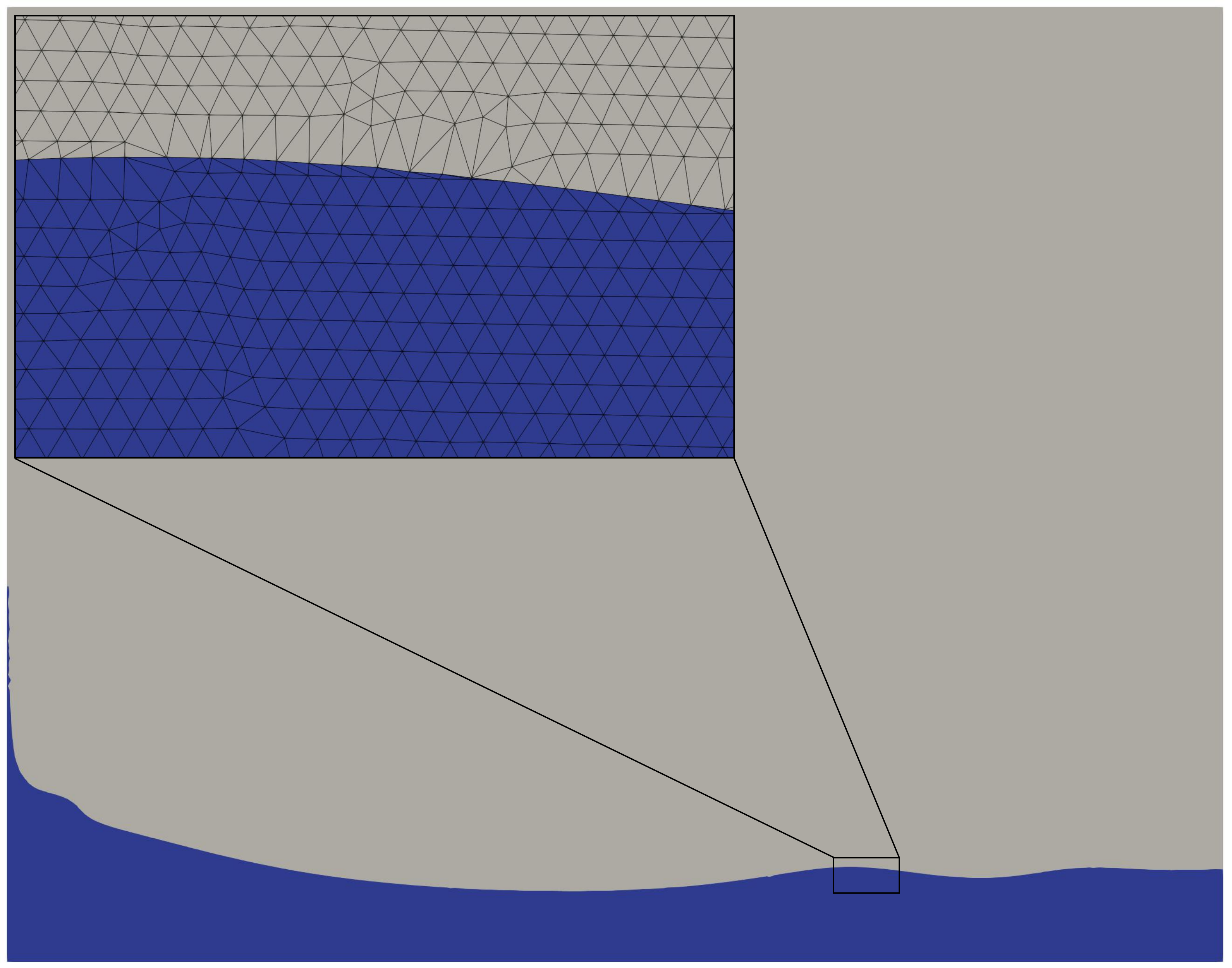}
    \end{subfigure}
    \caption{Dambreak simulation: Free surface position for the adimensional times $t =$ 0.5, 1.75, 3, 5.5, 6.5 and 22.5.} 
    \label{fig:dambreak}
\end{figure}

The mesh used has $75,950$ vertices. Figure \ref{fig:dambreak} represents the evolution of the free surface at the different adimensional times $t = 0.5, 1.75, 3, 5.5, 6.5$ and $22.5$, where the reference time considered is $t_{ref} = \sqrt{h/g}$. Despite the numerous topological changes present during a dambreak, the method is robust enough to simulate such complex flows.

\subsection{Rayleigh-Taylor}
 Rayleigh-Taylor instability consist in placing a heavier fluid above a lighter one. This position
is an unstable equilibrium and is sensible to any perturbation. A classical benchmark for the validation of two-phase flows solver is to trigger a single-mode perturbation in the initialisation. The instability develops under the influence of the gravity field and is commonly occurring in a wide range of physical phenomena. A disturbance is initialized in
the interface position at $t = 0$ and it’s initial position is given by: 
\begin{align*}
    y = 2.0 + 0.05 \cos{2\pi x}
\end{align*}
Initially, the velocity field is set to 0 and the pressure is hydrostatic. The evolution in time of the 2 fluids is determined by 2 adimensional numbers, the Atwood (At) and Reynolds (Re) numbers defined in our case by: 
\begin{align*}
    At &= \frac{\rho_1 - \rho_2}{\rho_1 + \rho_2} \\
    Re &= \frac{\sqrt{W g} W}{\nu}
\end{align*}
with $W$ the width of the canal.

\begin{figure}[!ht]
    \centering
    \captionsetup{font=normal}
    \begin{subfigure}[b]{0.16\textwidth}   
        \centering 
        \includegraphics[width=\textwidth]{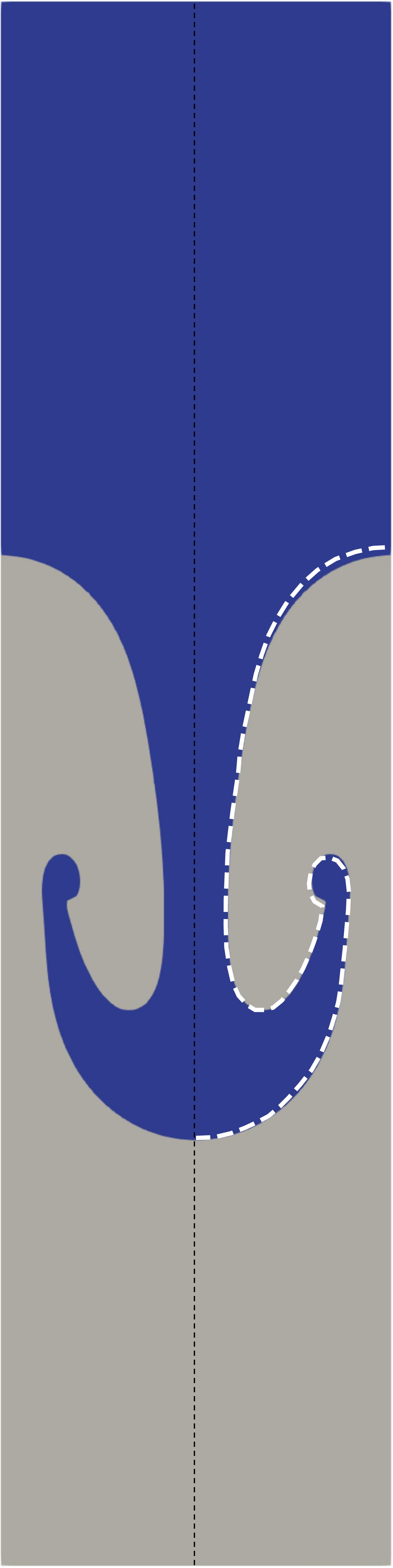}
    \end{subfigure}
    \hfill
    \begin{subfigure}[b]{0.16\textwidth}
        \centering
        \includegraphics[width=\textwidth]{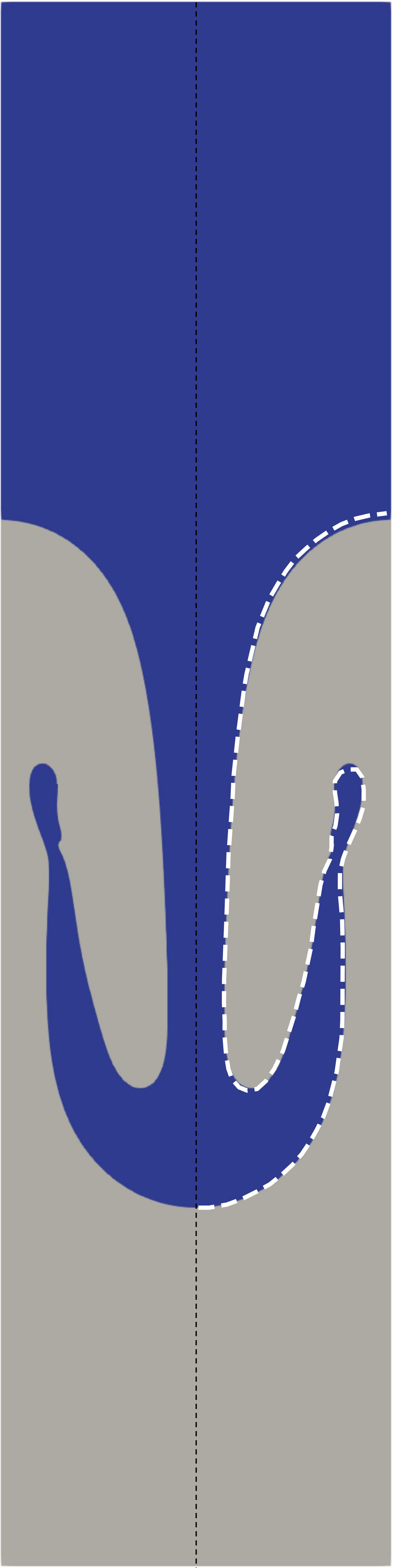}
    \end{subfigure}
    \hfill
    \begin{subfigure}[b]{0.16\textwidth}  
        \centering 
        \includegraphics[width=\textwidth]{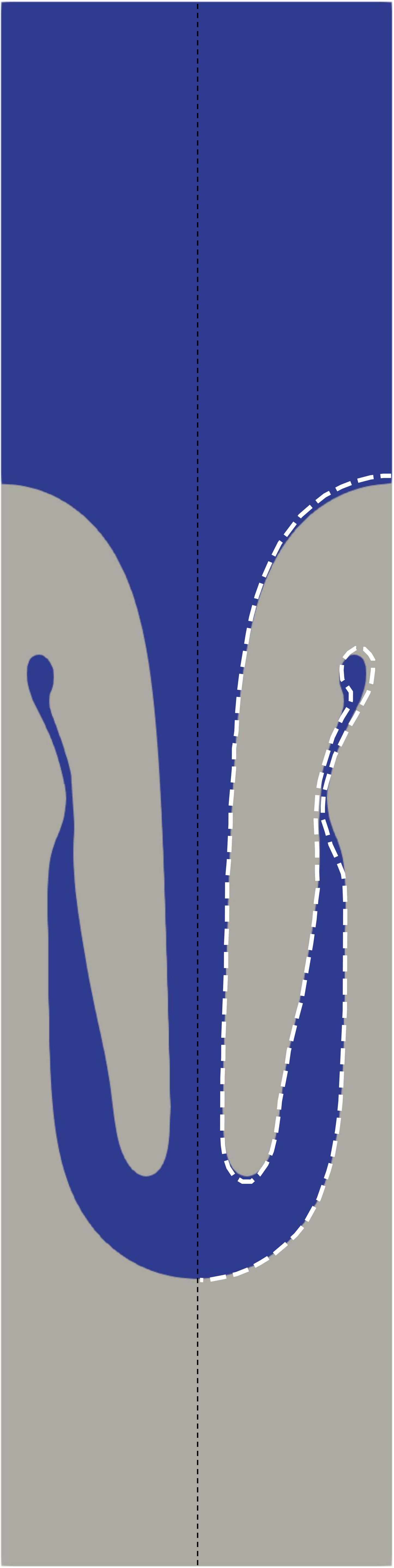}
    \end{subfigure}
    \hfill
    \begin{subfigure}[b]{0.16\textwidth}   
        \centering 
        \includegraphics[width=\textwidth]{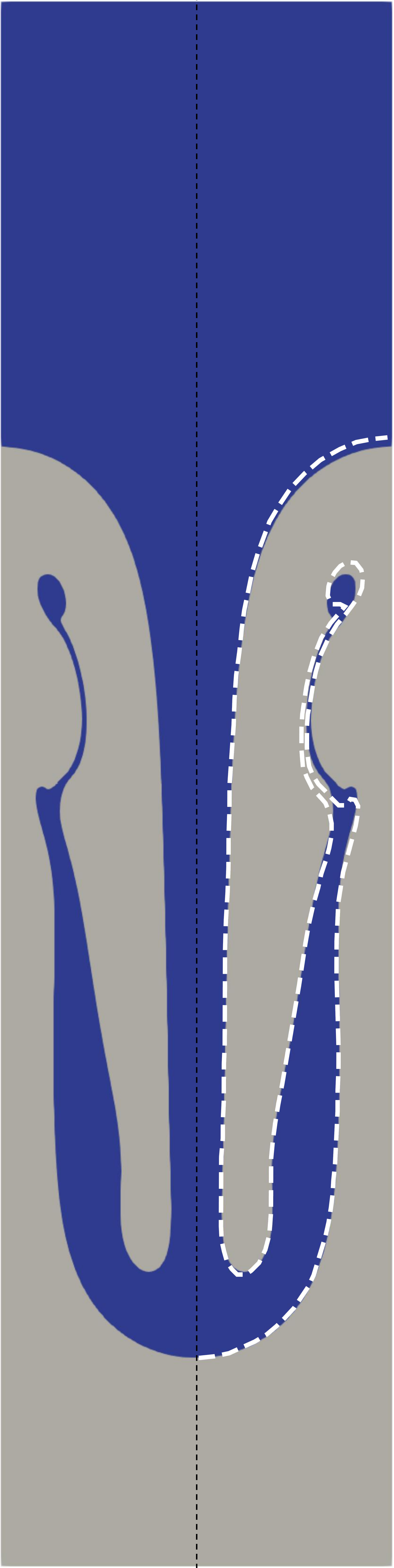}
    \end{subfigure}
    \begin{subfigure}[b]{0.16\textwidth}
        \centering
        \includegraphics[width=\textwidth]{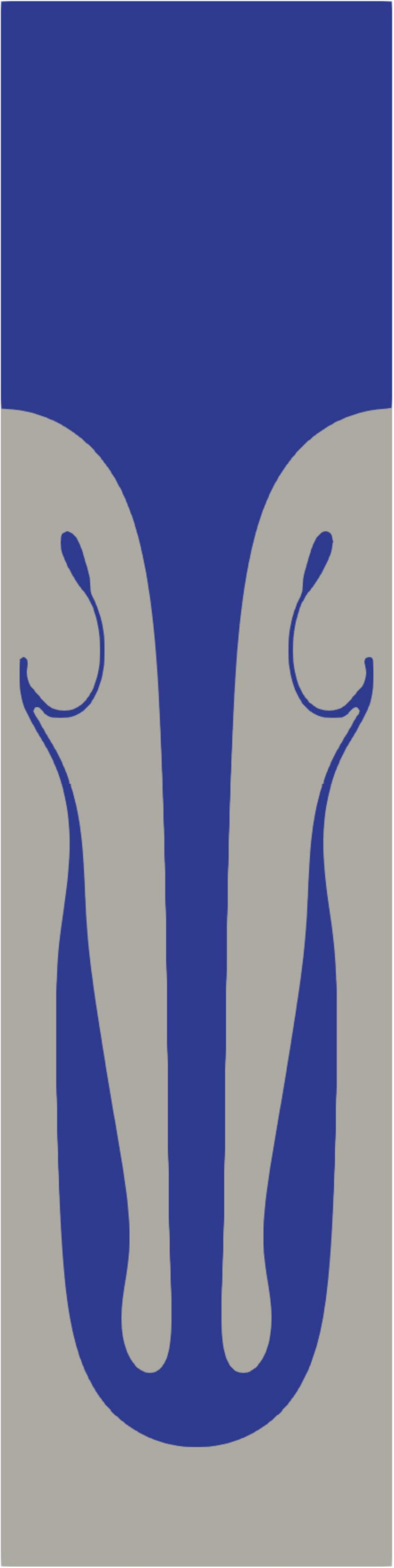}
    \end{subfigure}
    \hfill
    \begin{subfigure}[b]{0.16\textwidth}  
        \centering 
        \includegraphics[width=\textwidth]{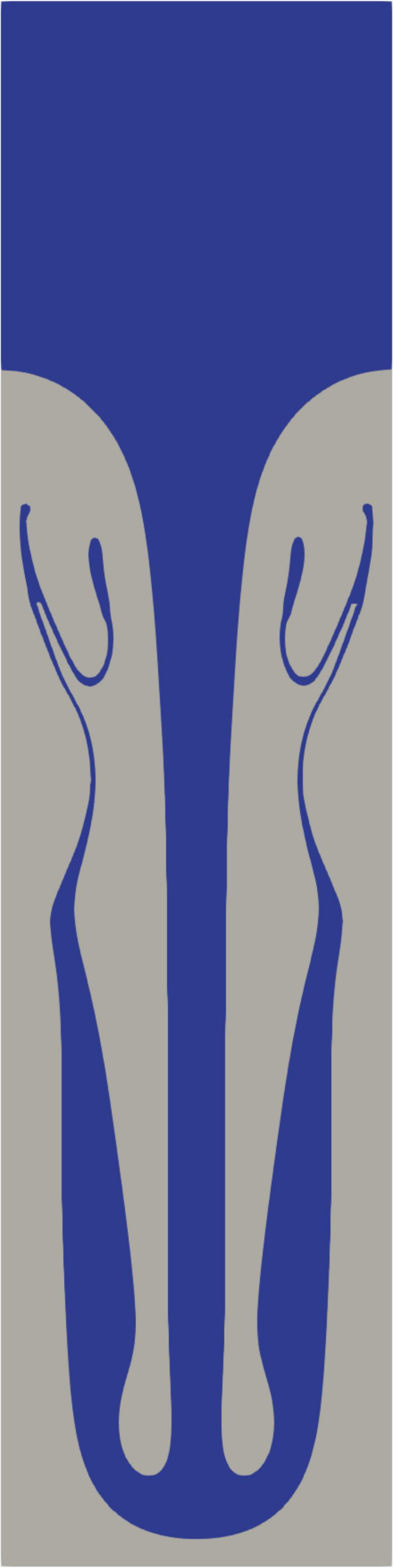}
    \end{subfigure}
    \hfill
    \\
    \begin{subfigure}[b]{0.16\textwidth}   
        \centering 
        \includegraphics[width=\textwidth]{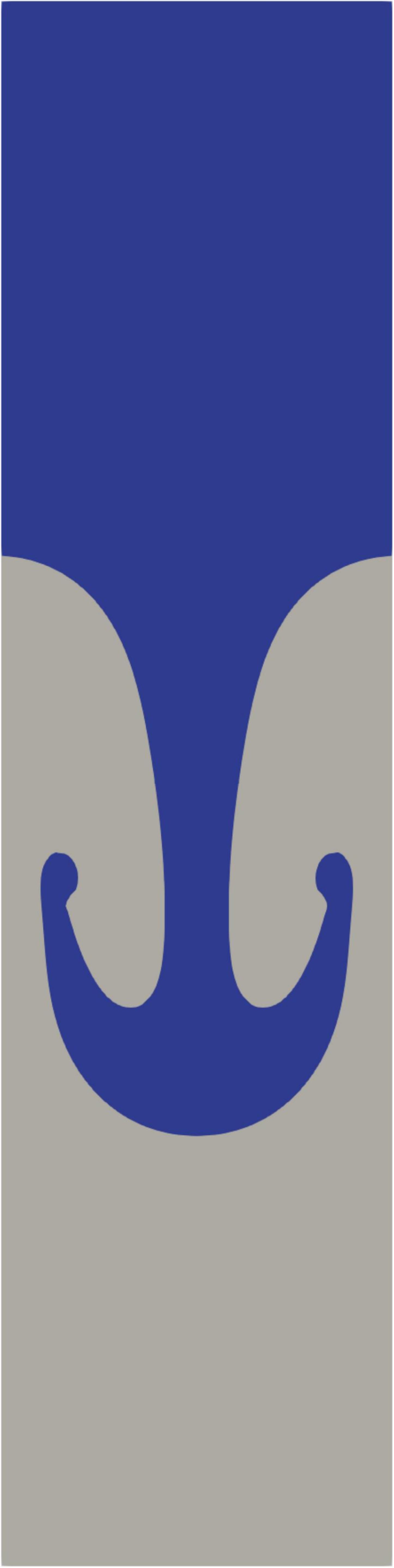}
    \end{subfigure}
    \hfill
    \begin{subfigure}[b]{0.16\textwidth}
        \centering
        \includegraphics[width=\textwidth]{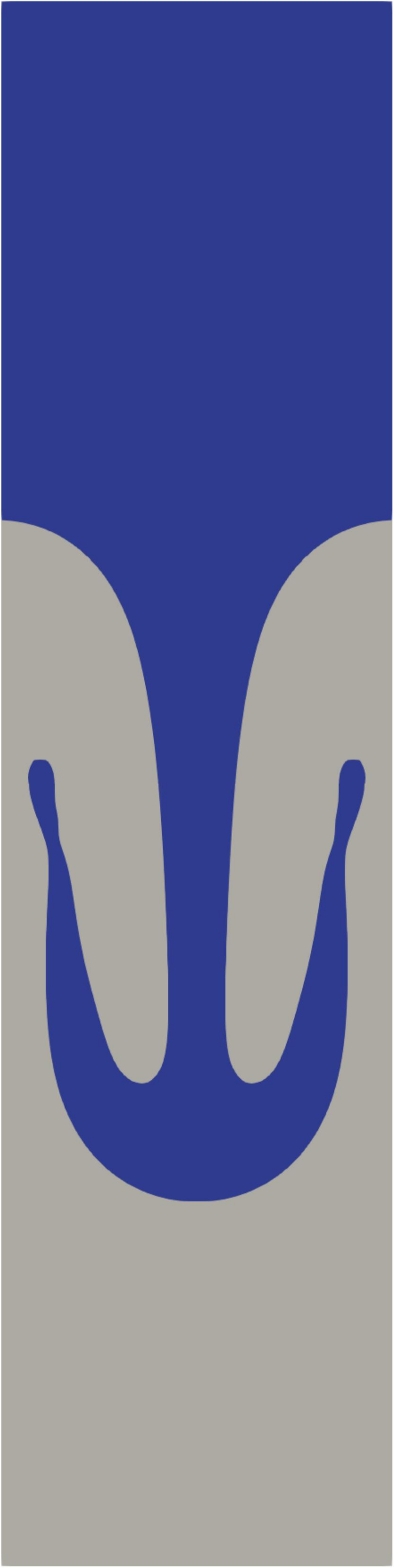}
    \end{subfigure}
    \hfill
    \begin{subfigure}[b]{0.16\textwidth}  
        \centering 
        \includegraphics[width=\textwidth]{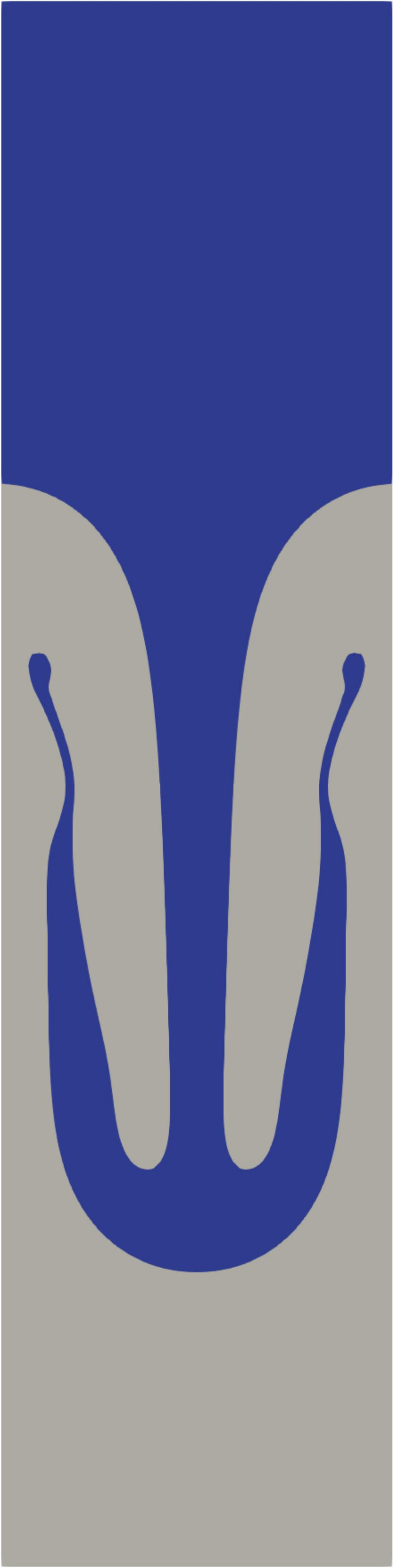}
    \end{subfigure}
    \hfill
    \begin{subfigure}[b]{0.16\textwidth}   
        \centering 
        \includegraphics[width=\textwidth]{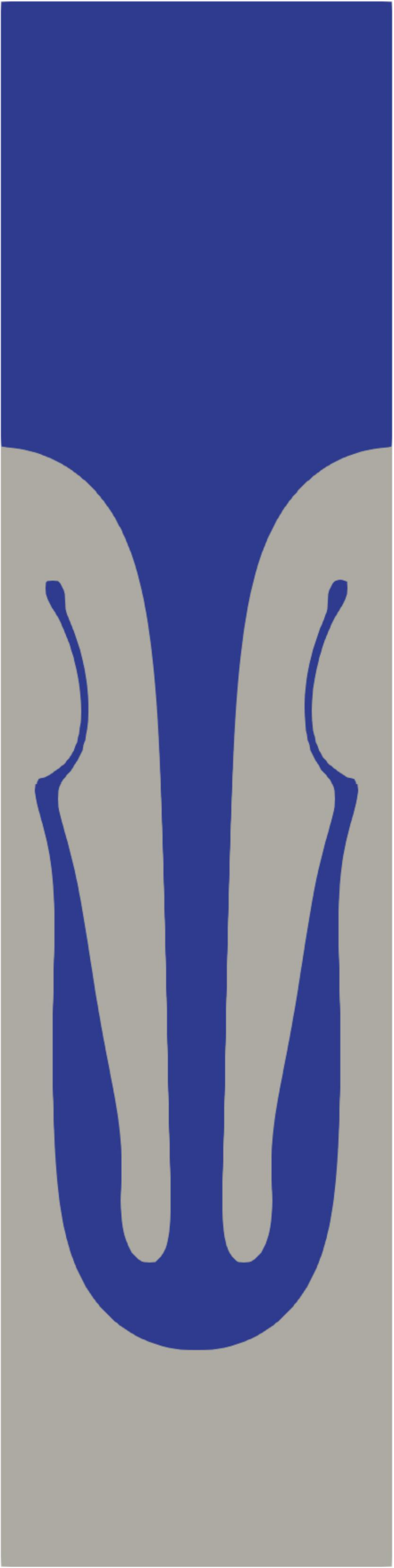}
    \end{subfigure}
    \begin{subfigure}[b]{0.16\textwidth}
        \centering
        \includegraphics[width=\textwidth]{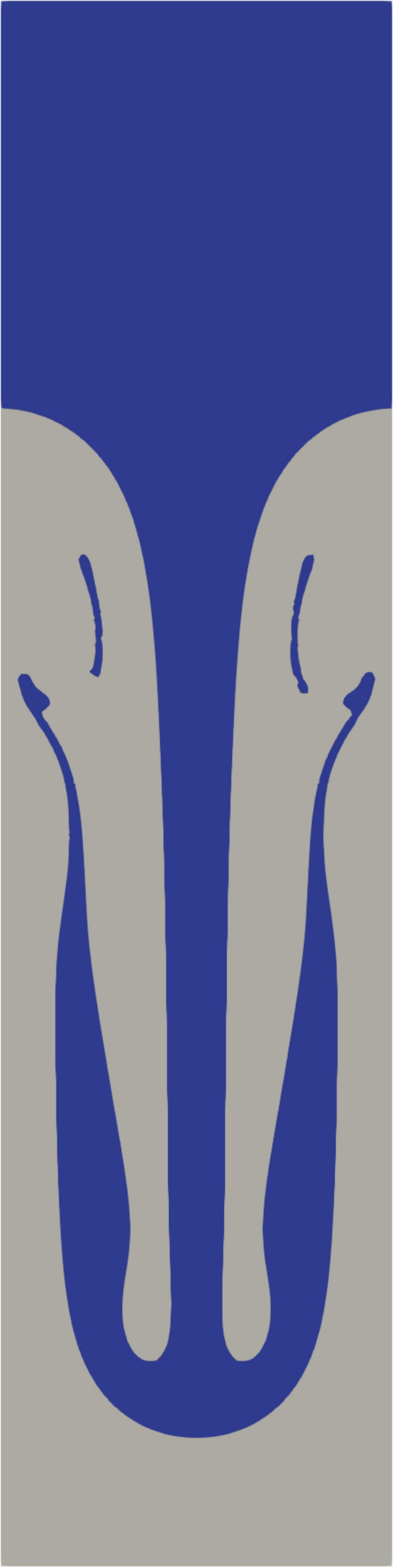}
    \end{subfigure}
    \hfill
    \begin{subfigure}[b]{0.16\textwidth}  
        \centering 
        \includegraphics[width=\textwidth]{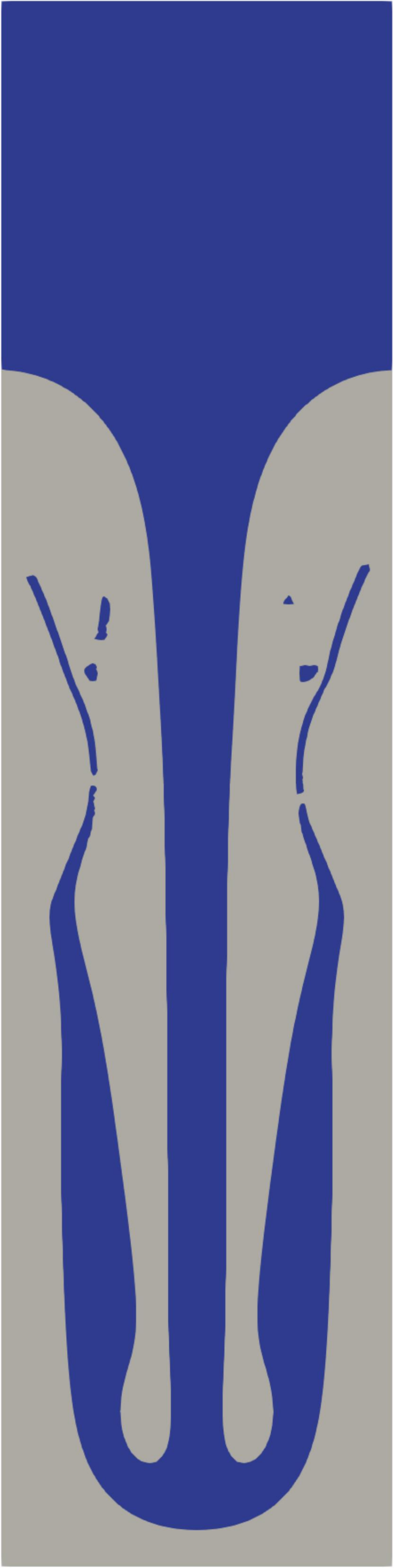}
    \end{subfigure}
    \caption{Rayleigh-Taylor instability for $Re = 256$, $At = 0.5$ at times  $t = 3, 3.5, 4, 4.5, 5, 5.5$ with a fine mesh (top) and a coarser mesh (bottom).} 
    \label{fig:RT256}
\end{figure}

\begin{figure}[!ht]
    \centering
    \captionsetup{font=normal}
    \begin{subfigure}[b]{0.16\textwidth}   
        \centering 
        \includegraphics[width=\textwidth]{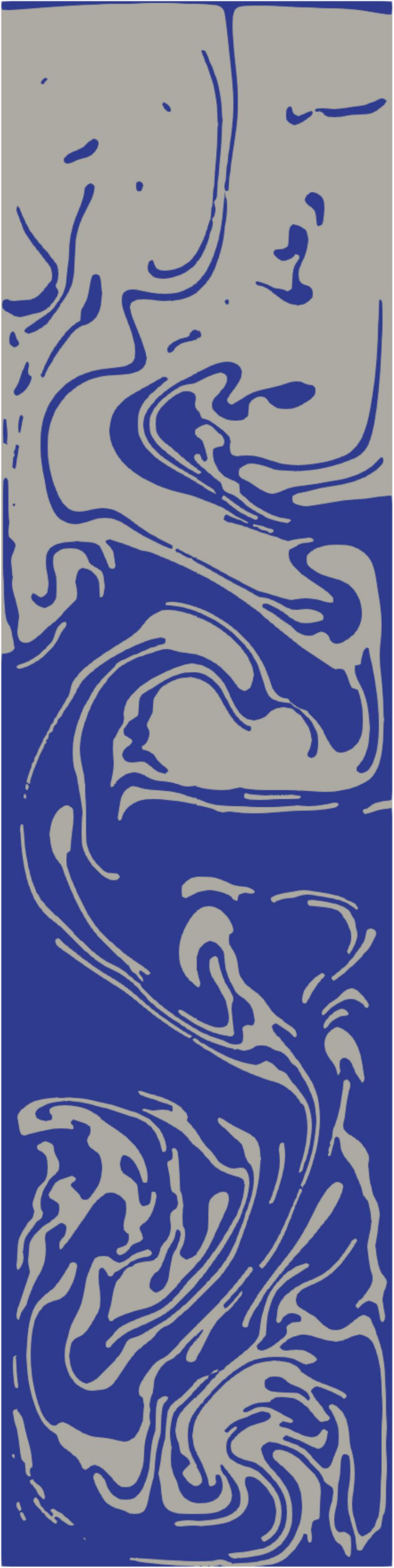}
    \end{subfigure}
    \begin{subfigure}[b]{0.75\textwidth}   
        \centering 
        \includegraphics[width=\textwidth]{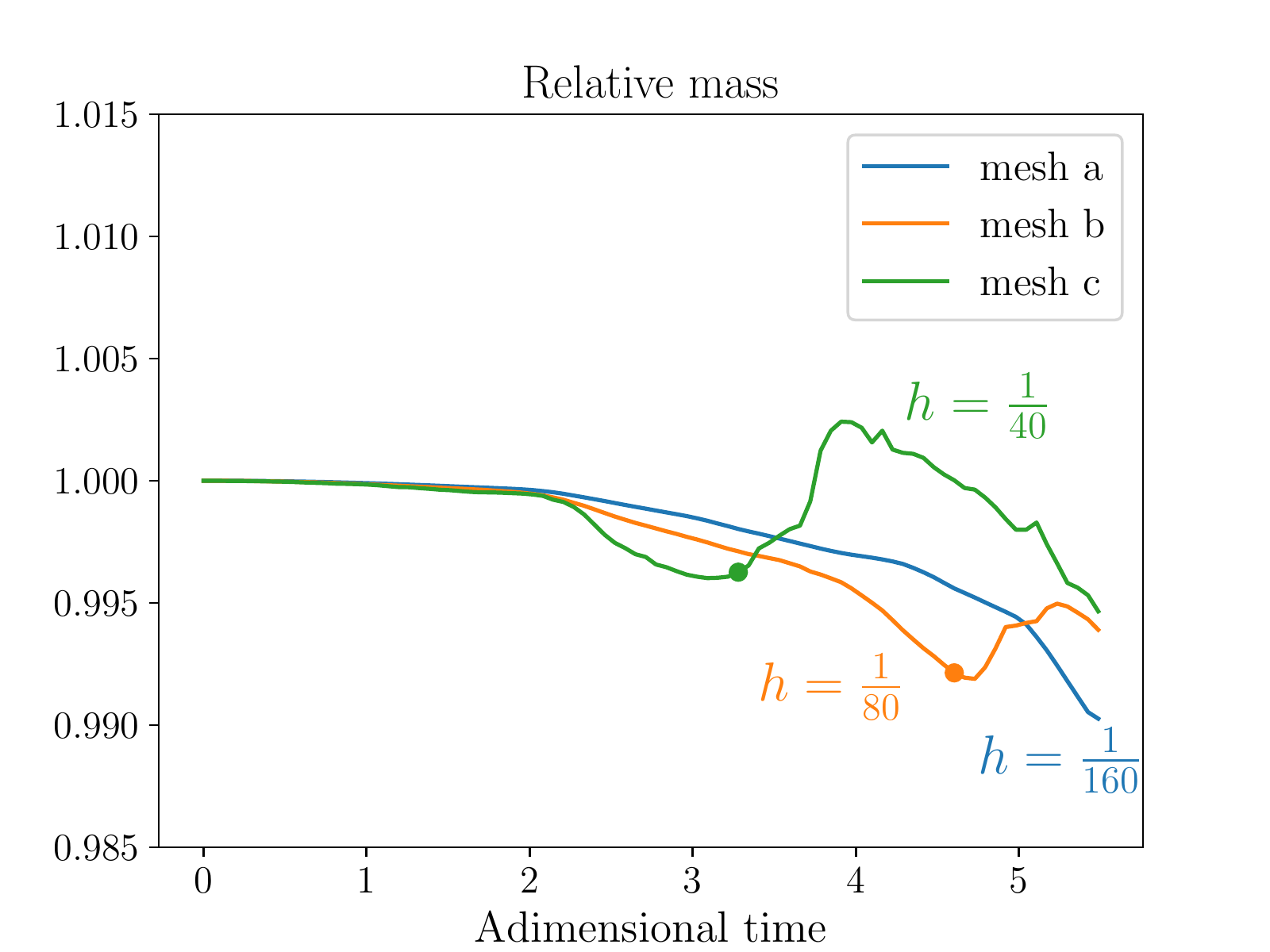}
    \end{subfigure}
    \hfill
    \caption{(left) Rayleigh-Taylor instability at time $t=15$ with the mesh a . (right) Relative mass of the grey fluid during the Rayleigh-Taylor simulation for the 3 different meshes}
    \label{fig:RT15_mass}
\end{figure}

Figure \ref{fig:RT256} shows the evolution of the two phases for the case At = 0.5 and Re = 256.   We ran our simulations on three different meshes: mesh a with approximately 123 000 nodes, mesh b with 31 000 nodes and a third very coarse mesh c with 8250 nodes. We compared our results with those of He et al [21] for a mesh of approximately 262,000 elements using the lattice Boltzmann method. The dashed white line in Figure 1 correspond to the interface they obtained. The sharp interface of the method presented, as well as the fine mesh 1 used, allows the features of the flow to be preserved for a long time. As explained in section 3.3.3, if the 3 nodes of an element are positioned on the interface, the phase of this element is ambiguous. This happens every time a phase merges or splits because the level set naturally handles topological changes at the scale of an element. The process of merging is thus mesh dependent and the asymmetry of the mesh generates a small asymmetry in the flow which propagates until it generates a completely asymmetric flow, as we can see in Figure \ref{fig:RT15_mass} (left). The effect of the characteristic length of the mesh is shown in shown in Figure \ref{fig:RT256} (bottom), where the same Rayleigh-Taylor instability has been simulated but with the 4 times coarser mesh b. An analytical solution for the linearized equations exist and is valid for the linear phase of the instability development \cite{drazin2004hydrodynamic}:
\begin{align*}
    h = h_0 e^{\hat{\alpha}t}
\end{align*}
with $h$ the perturbation size, $h_0$ the perturbation at $t=0$ and $\hat{\alpha}$ the growth rate. For the parameter of the simulation the analytical grow rate is $\hat{\alpha}_{a} = 9.3$ and the observed one is $\hat{\alpha}_{o} = 9.2$. 

 It is also interesting to compare the mass conservation for different mesh refinements. The evolution of the relative mass of the grey fluid during the Rayleigh-Taylor simulation is shown in Figure \ref{fig:RT15_mass}. We can see that the mass variation is limited to 1\% and is more chaotic for coarser meshes. The dots on the graph represent the moment when the flow is no longer completely captured by the mesh. As we can see the mass start to strongly vary once that point is reached. By placing the nodes of the mesh on the boundary, we naturally \textit{cut off the corners} of the boundary, resulting in a loss of mass for Fluid 1 or Fluid 2, depending on whether the boundary is convex or concave. The Rayleigh-Taylor instability has an interface that is convex in some places and concave in others. This leads to opposite effects of mass loss, which explain the ascending and descending aspects of the curves in Figure \ref{fig:RT15_mass}.

\subsection{Single Bubble rising}
Bubble dynamics are frequently encountered in a variety of industrial processes or natural flows. The correct representation of the interface is essential for this type of problem because the surface tension force is proportional to the curvature of the interface. Grace et al \cite{grace} have characterized in 1976 the final shape obtained during single bubble rise experiment for different fluids. Figure \ref{fig:grace} summarizes their observations. To validate our method for bubble dynamics we try to reproduce theses results. The y axis of Figure \ref{fig:grace} correspond to the Reynolds number and it's x axis to the Bond (or Eötvos) number. Theses number can be computed for the case of bubbles by \begin{align*}
    Re &= \frac{\rho_1 \sqrt{g} D^{3/2}}{\mu_1}\\
    Bo &= \frac{\rho_1 g D^2}{\sigma}
\end{align*} 
where $D$ is the diameter of the bubble and the subscripts $1$ corresponds to the heavier fluid.

\begin{figure}[!ht]
    \centering
    \captionsetup{font=normal}
    \begin{subfigure}[b]{0.29\textwidth}
        \centering
        \includegraphics[width=\textwidth]{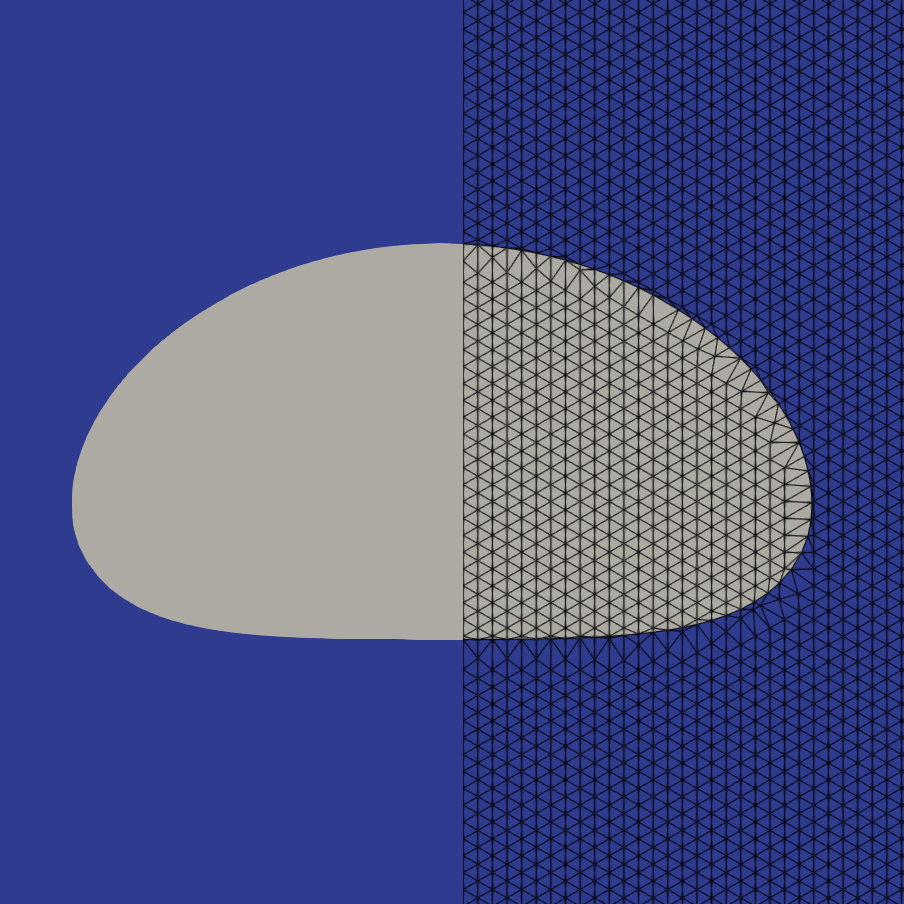}
        {(a) $Re = 10$, $Bo = 10$}
    \end{subfigure}
    \hfill
    \begin{subfigure}[b]{0.29\textwidth}  
        \centering 
        \includegraphics[width=\textwidth]{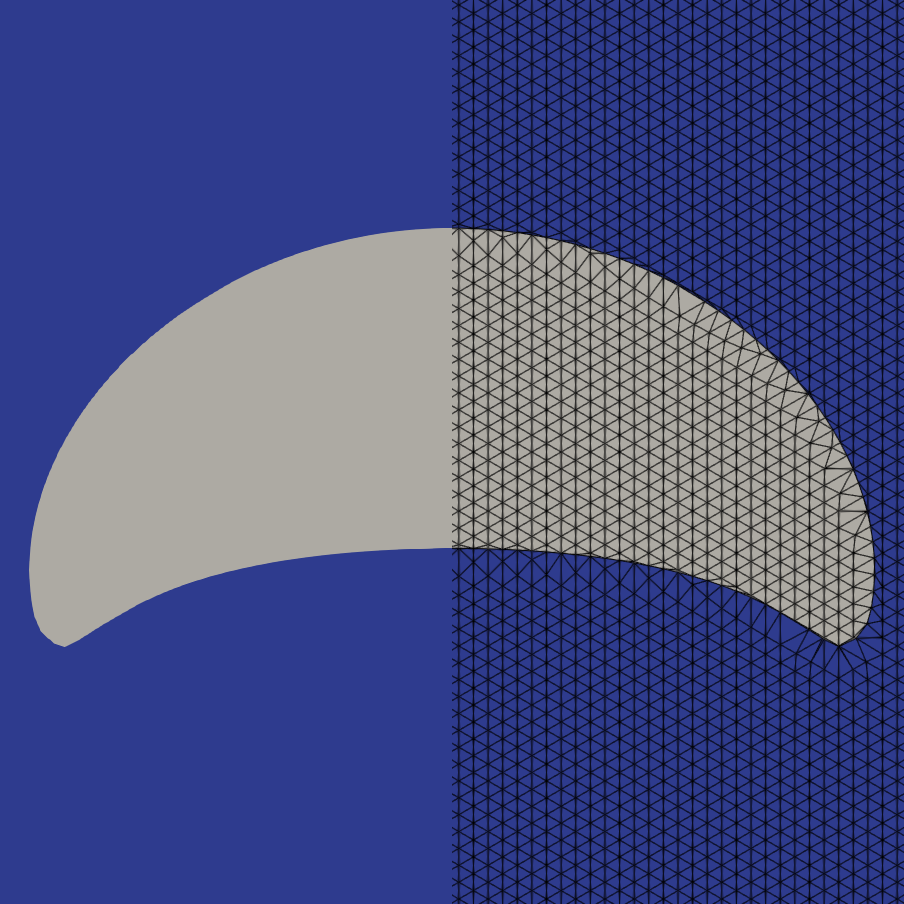}
        {(b) $Re = 10$, $Bo = 50$}
    \end{subfigure}
    \hfill
    \begin{subfigure}[b]{0.36\textwidth}  
        \centering 
        \includegraphics[width=\textwidth]{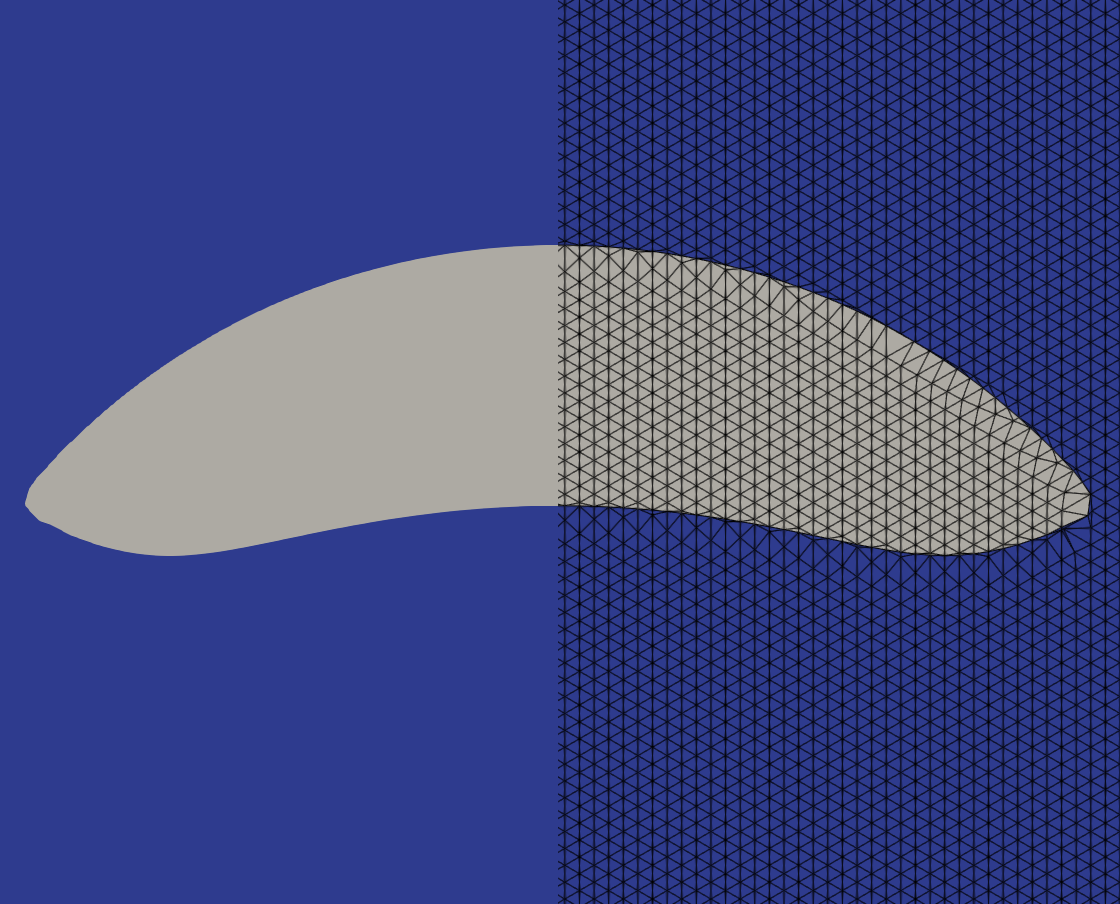}
        {(c) $Re = 100$, $Bo = 50$}
    \end{subfigure}
    \caption{Final form of free rise bubbles}
    \label{fig:freerise}
\end{figure}

\begin{figure}[!ht]
    \centering
    \captionsetup{font=normal}
    \includegraphics[width=0.5\textwidth]{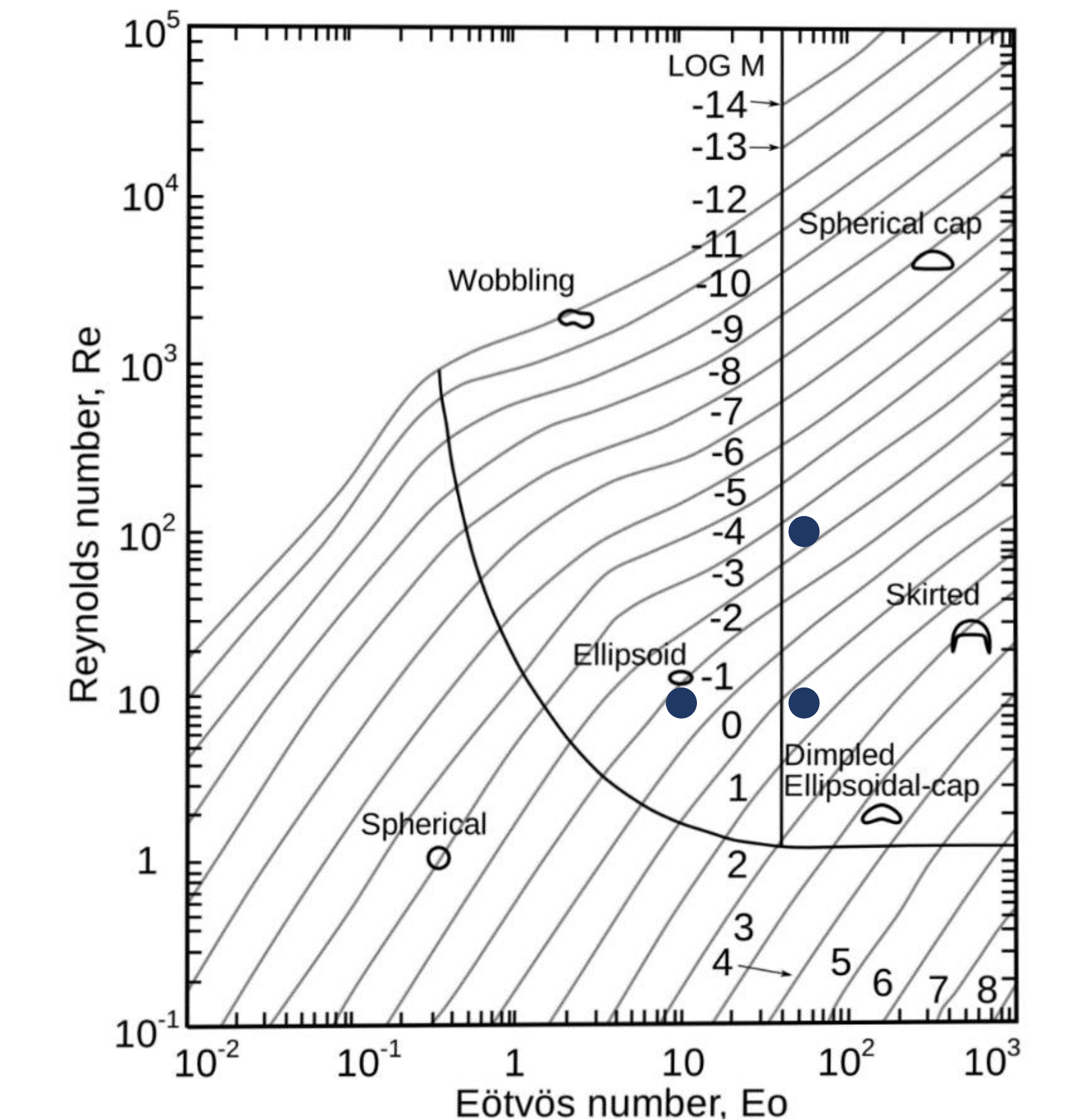}
    \caption{Grace's diagram \cite{grace}.}
    \label{fig:grace}
\end{figure}

We consider three cases with different pairs of Reynolds and Bond numbers represented in Figure \ref{fig:freerise} and on Grace's diagram at Figure \ref{fig:grace}. The density and viscosity ratios considered are $\frac{\rho_1}{\rho2} = 1000$ and $\frac{\mu_1}{\mu_2} = 100$. The diameter of the bubble is fixed to $D=1 \hspace{2pt} [m]$ and to avoid any impact of the lateral wall on the bubble dynamic the computational domain considered is $W=6 \hspace{2pt} [m]$ width and $H = 12 \hspace{2pt} [m]$ high. The surface tension is applied explicitly engendering a stability condition on the time step. The time step must verify the condition:
\begin{align*}
    \Delta t < \sqrt{\frac{\left( \rho_1 + \rho_2 \right) h^3}{ 4 \pi \sigma}}
\end{align*}
where $h$ is the characteristic size of the mesh.

When we look at the final shape of the bubbles and compare it with Grace's experimental diagram or with Hua \textit{et al} \cite{hua2007numerical}, we see that the simulations follow the same trends. A good way of validating our results is to compare them using the 2D benchmark proposed by Hygwing et al. This is a free rising bubble with the following properties for the 2 fluids:
\begin{align*}
    \rho_1 = 1000 \hspace{1cm} \rho_2=100 \hspace{1cm} \mu_1 = 10 \hspace{1cm} \mu_2=1 \hspace{1cm} g=0.98 \hspace{1cm} \sigma = 24.5
\end{align*}
In addition to comparing the final shape of the bubbles, the paper proposes to compare several benchmark quantities: the centre of mass of the bubble, the mean velocity of the bubble and the circularity. The latter is defined as the ratio between the perimeter of a circle of the same area as the bubble and the perimeter of the bubble. We will also add the mass (area) of the bubble. To observe the convergence of the method, this simulation is carried out with different levels of mesh refinement: $h= \frac{1}{40}, \frac{1}{80} \text{ and } \frac{1}{160}$ where $h$ is the characteristic size of an element. A reference solution is also displayed, the solution obtained by the TP2D solver \cite{OSHER198812, turek1997discrete} for a mesh size $h=\frac{1}{320}$. TP2D, which is short for Transport Phenomena in 2D, is a code developped by TU Dortmund and is an extension of the Featflow incompressible flow solver to treat immiscible fluids with the level set method.
\begin{figure}[!ht]
    \centering
    \captionsetup{font=normal}
    \begin{subfigure}[b]{0.45\textwidth}
        \centering
        \includegraphics[width=\textwidth]{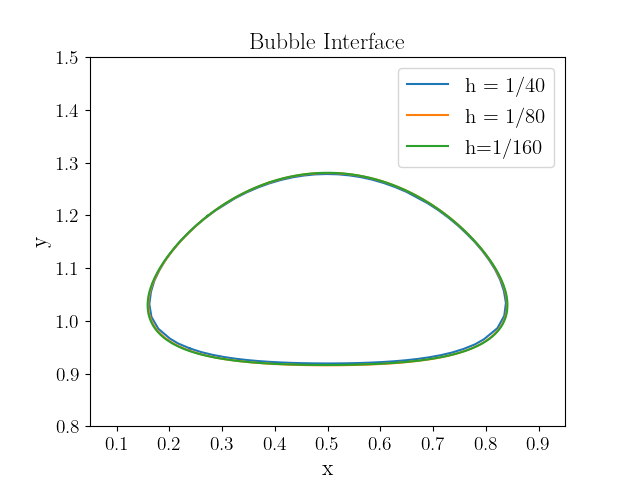}
    \end{subfigure}
    \hfill
    \begin{subfigure}[b]{0.45\textwidth}  
        \centering 
        \includegraphics[width=\textwidth]{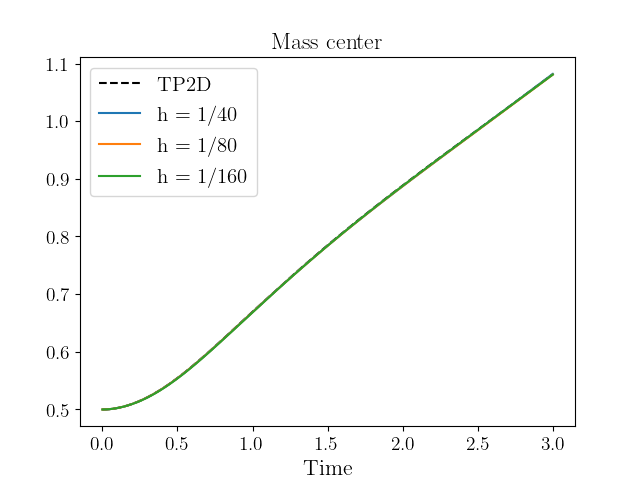}
    \end{subfigure}
    \\
    \vspace{0.2cm}
    \begin{subfigure}[b]{0.45\textwidth}
        \centering
        \includegraphics[width=\textwidth]{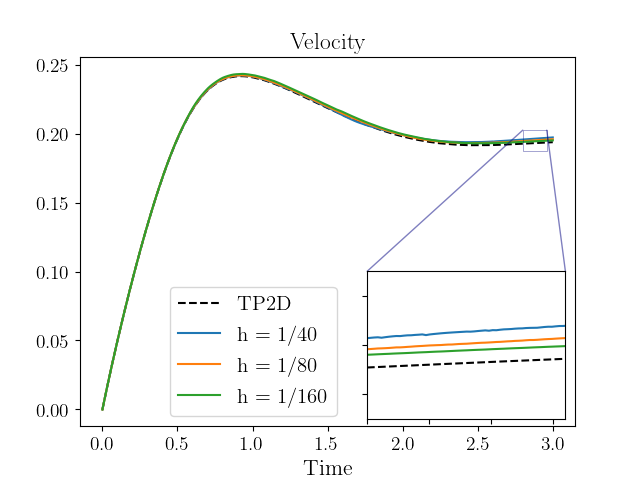}
    \end{subfigure}
    \hfill
    \begin{subfigure}[b]{0.45\textwidth}  
        \centering 
        \includegraphics[width=\textwidth]{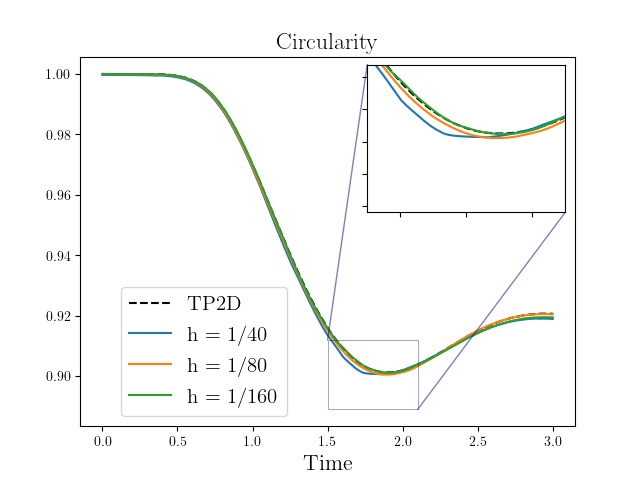}
    \end{subfigure}
    \\
    \vspace{0.2cm}
    \centering
    \begin{subfigure}[b]{0.45\textwidth}
        \centering
        \includegraphics[width=\textwidth]{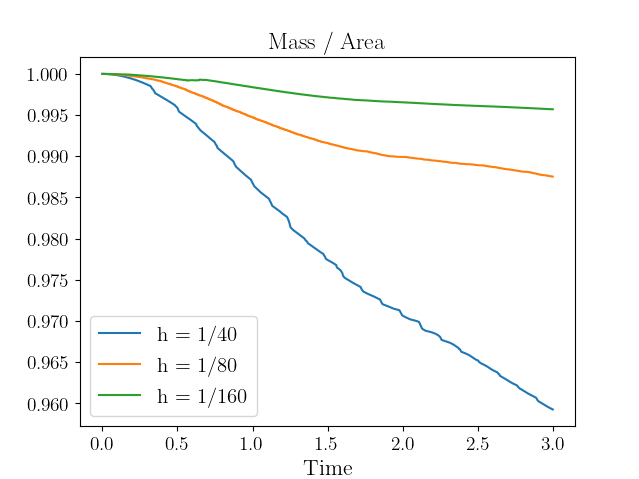}
    \end{subfigure}
    \caption{Results of the benchmark of Hysing et al. for a free rising bubble.}
    \label{fig:hysing}
\end{figure}

As we can see on Figure \ref{fig:hysing}, the bubble interface converge quite quickly as the interface position for $h=\frac{1}{80}$ and $h=\frac{1}{160}$ are difficult to distinguish. The mass center vertical position of the bubble as well as the velocity and the circularity of the bubble can be compared with the reference solution. The position of the mass center is in good agreement with the reference and the solution obtained with our method is close to the reference solution even for a coarse mesh. The velocity of the bubble, measured as the integral of the velocity inside the bubble, doesn't seem to converge but the solution is already almost on the reference solution. The circularity is a more complex benchmark quantity to obtain in the simulations. We observe in Figure \ref{fig:hysing} that our method converge to the reference solution but there is still an important difference. Finally we can compute the relative mass of the bubble evolving during the simulation. We can clearly see the impact of the mesh on the mass conservation with a variation going from approximately 4 \% for a characteristic mesh size $h=\frac{1}{40}$ to 0.25 \% for $h=\frac{1}{160}$.

\subsection{Bubble merging}
To verify that the approach used for the surface tension works during topological changes of the fluid phases, we are interested in a test case of two bubbles superimposed in a column of heavy fluid with a free surface. The computational domain is  a 6 by 3 $[m]$ square, the two bubbles are centered horizontally, their center are positionned at height $y=1 \hspace{2pt} [m]$ and $y=2 \hspace{2pt} [m]$ with radius of 0.4 and 0.5 meters respectively. The parameters of the two fluids are chosen such that:
\begin{align*}
    \frac{\rho_1}{\rho_0} = 1000 \hspace{1.5cm}
    \frac{\mu_1}{\mu_0} = 2 \hspace{1.5cm}
    Re = 104  \hspace{1.5cm}
    Bo = 313\\
\end{align*}
with the $Re$ and $Bo$ numbers defined with the diameter of the greater bubble.

The lower bubble is sucked by the drag of the upper bubble such that at $t=0.4 \hspace{2pt} [s]$ it is engulfed by the latter. The two bubbles rise together until at $t=0.95\hspace{2pt} [s]$, the stronger currents in the center of the column push them to collapse. This leads to multiple changes in topology, resulting in two bubbles that rise to the free surface and two elongated bubbles that are carried to the bottom by downward currents on the sides. 

\begin{figure}[!ht]
    \centering
    \captionsetup{font=normal}
    \begin{subfigure}[b]{0.24\textwidth}
        \centering
        \includegraphics[width=\textwidth]{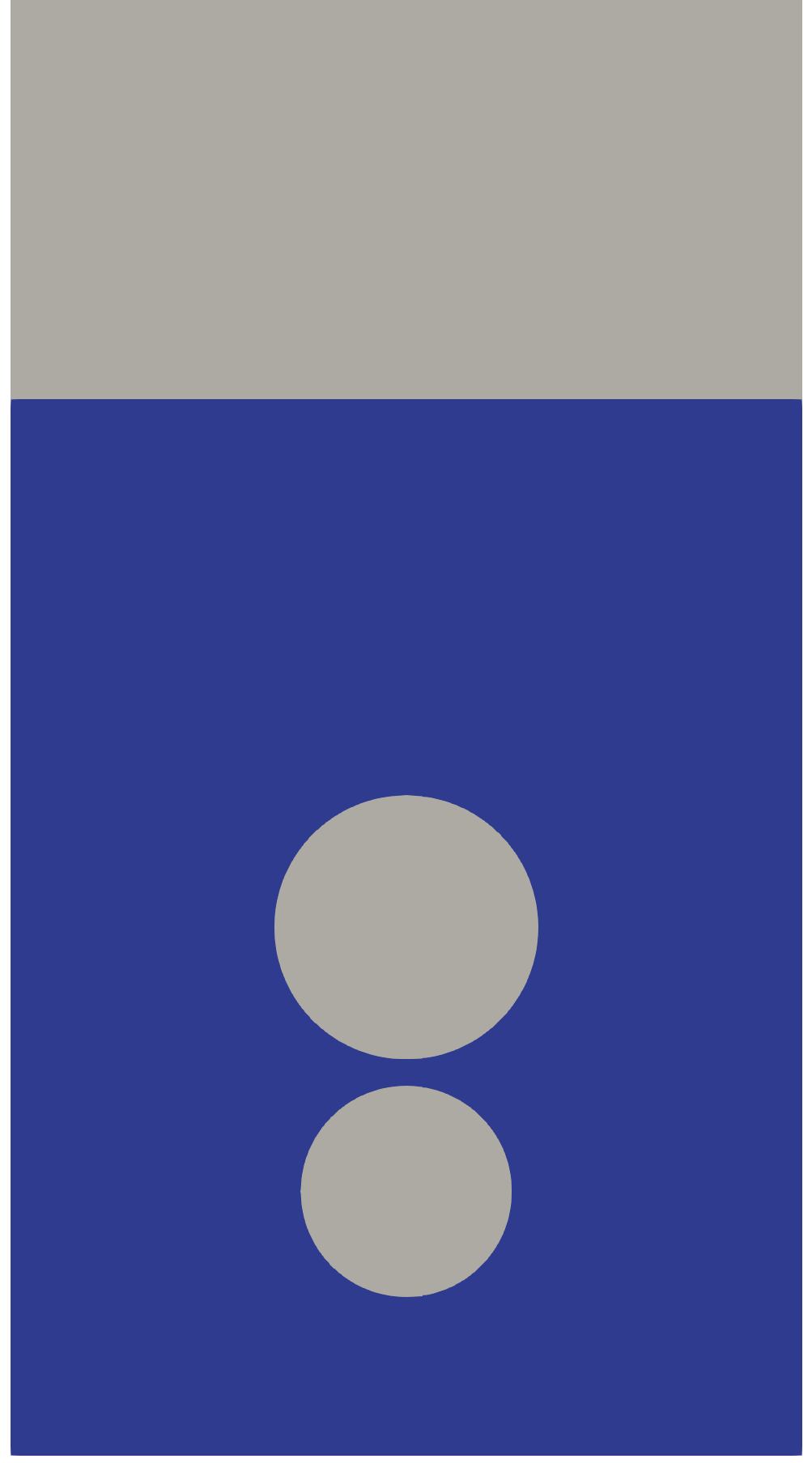}
        \caption{$t=0$}
    \end{subfigure}
    \hfill
    \begin{subfigure}[b]{0.24\textwidth}  
        \centering 
        \includegraphics[width=\textwidth]{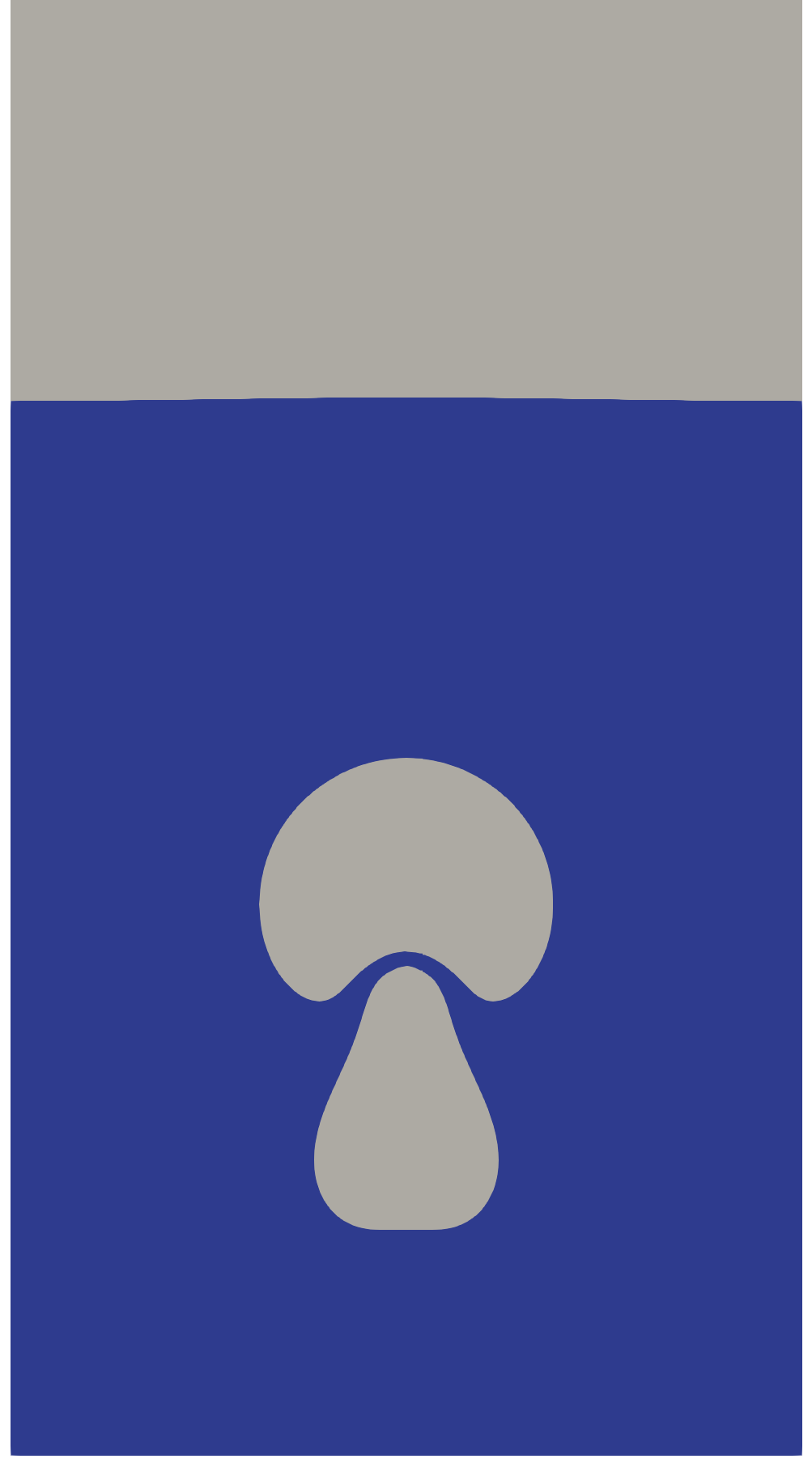}
        \caption{$t=0.2$}
    \end{subfigure}
    \hfill
    \begin{subfigure}[b]{0.24\textwidth}  
        \centering 
        \includegraphics[width=\textwidth]{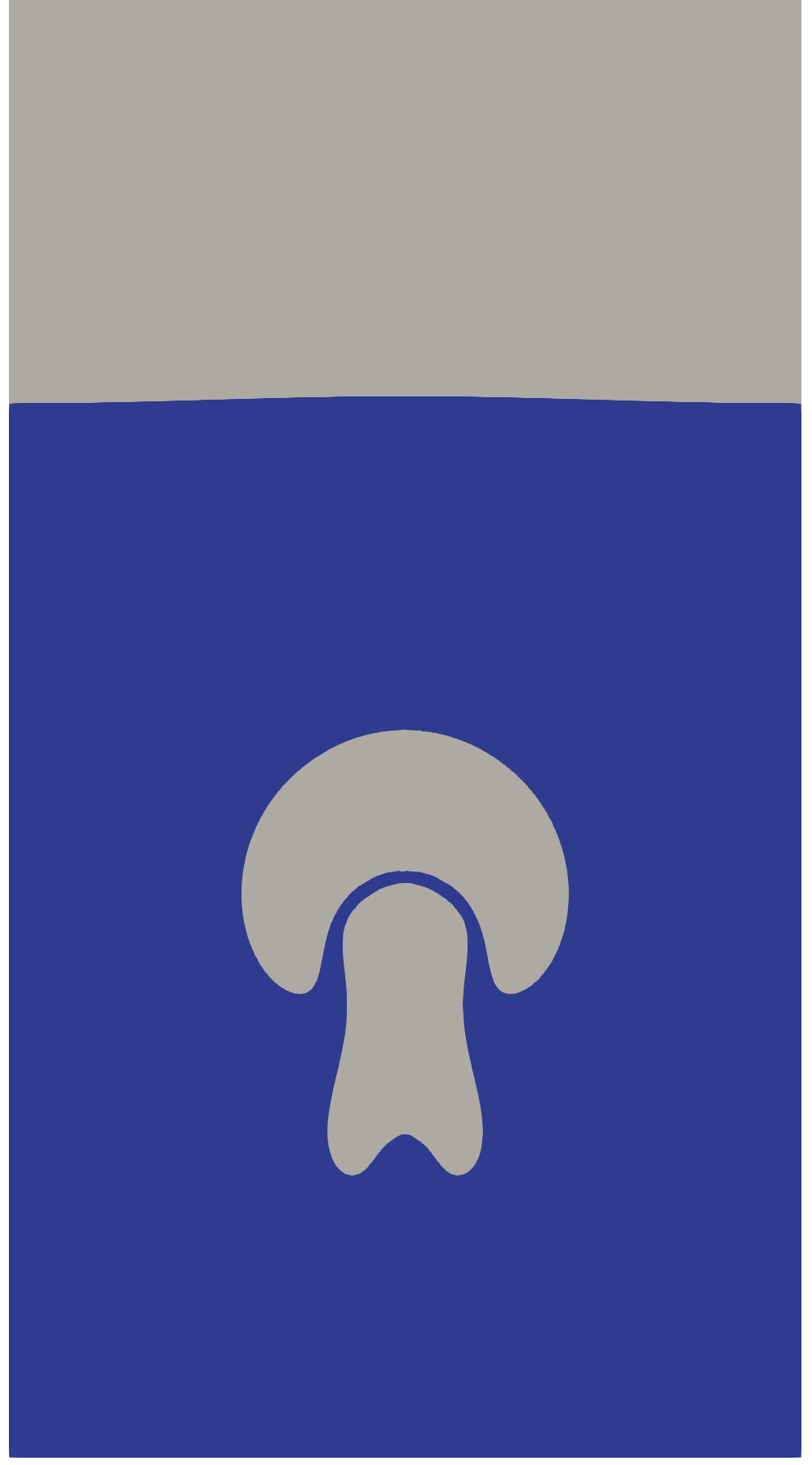}
        \caption{$t=0.3$}
    \end{subfigure}
    \hfill
    \begin{subfigure}[b]{0.24\textwidth}  
        \centering 
        \includegraphics[width=\textwidth]{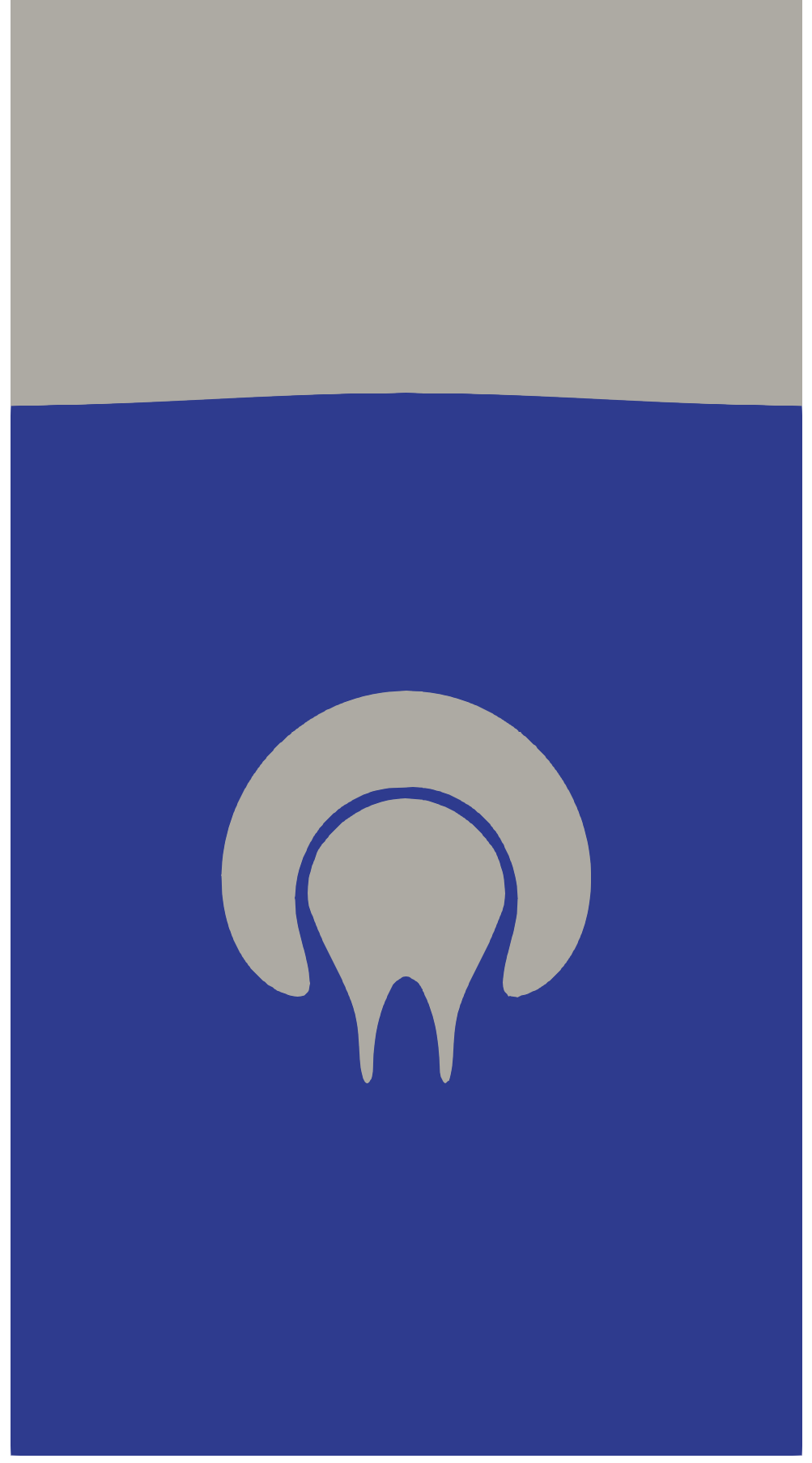}
        \caption{$t=0.4$}
    \end{subfigure}
    \\
    \centering
    \begin{subfigure}[b]{0.24\textwidth}
        \centering
        \includegraphics[width=\textwidth]{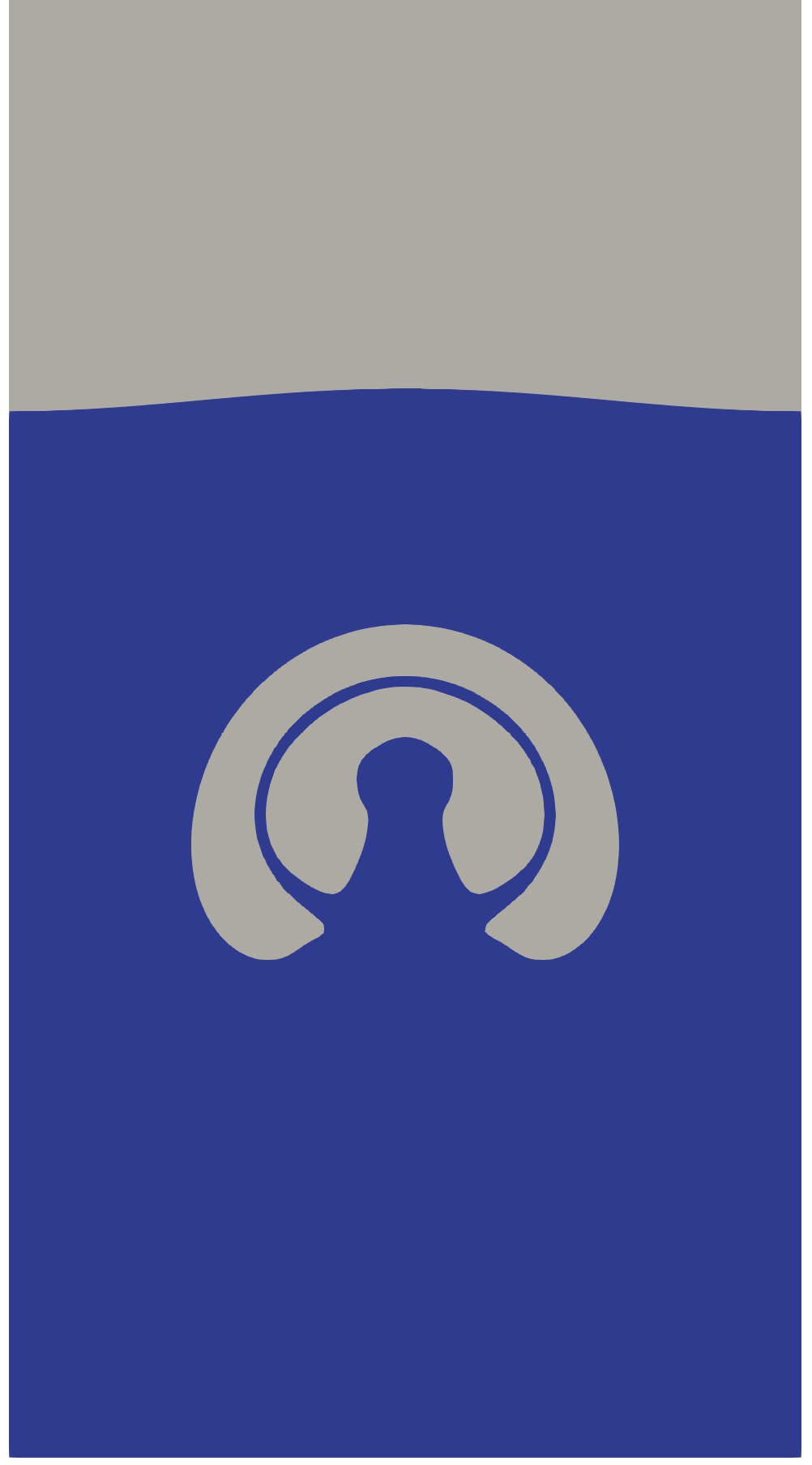}
        \caption{$t=0.6$}
    \end{subfigure}
    \hfill
    \begin{subfigure}[b]{0.24\textwidth}  
        \centering 
        \includegraphics[width=\textwidth]{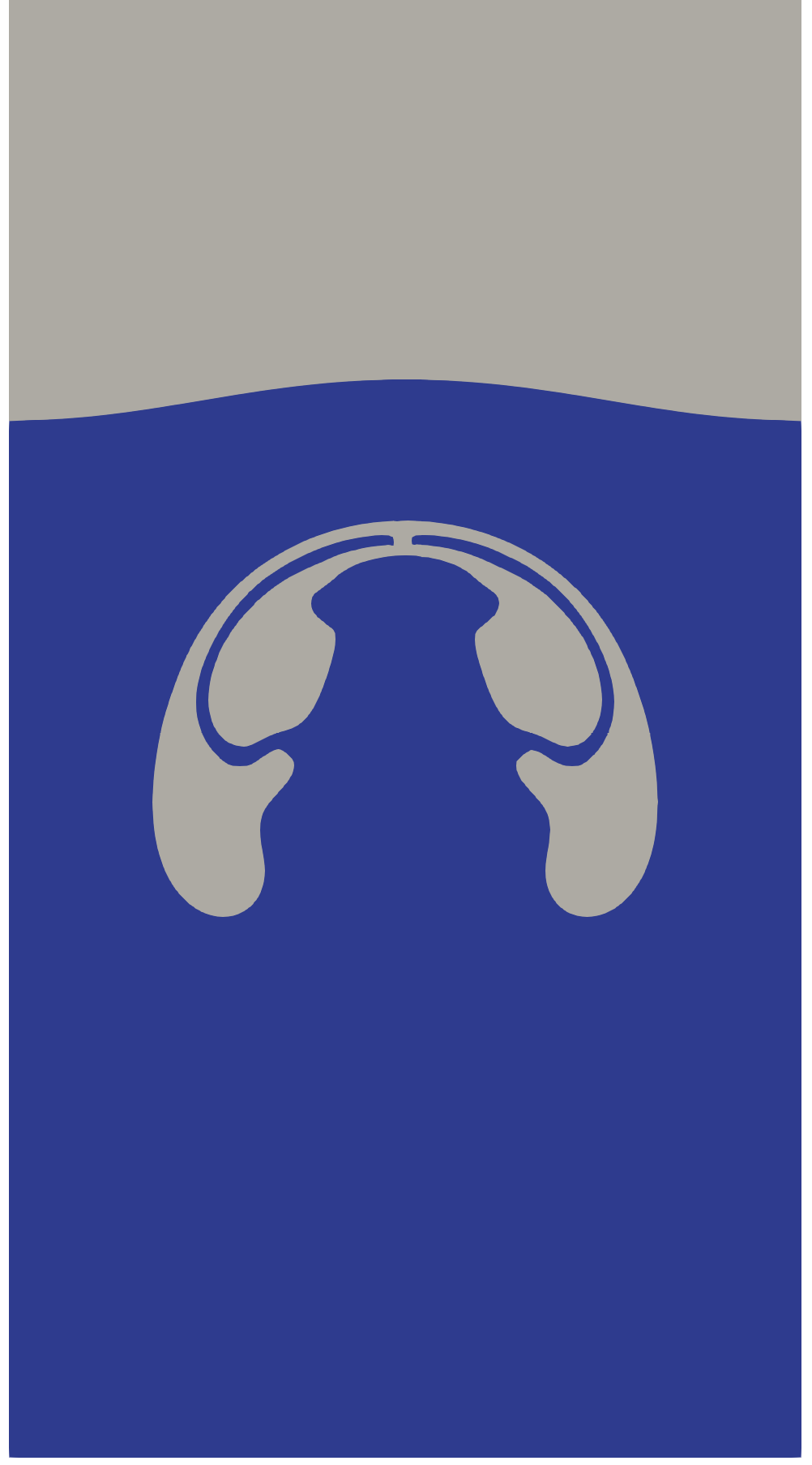}
        \caption{$t=0.95$}
    \end{subfigure}
    \hfill
    \begin{subfigure}[b]{0.24\textwidth}  
        \centering 
        \includegraphics[width=\textwidth]{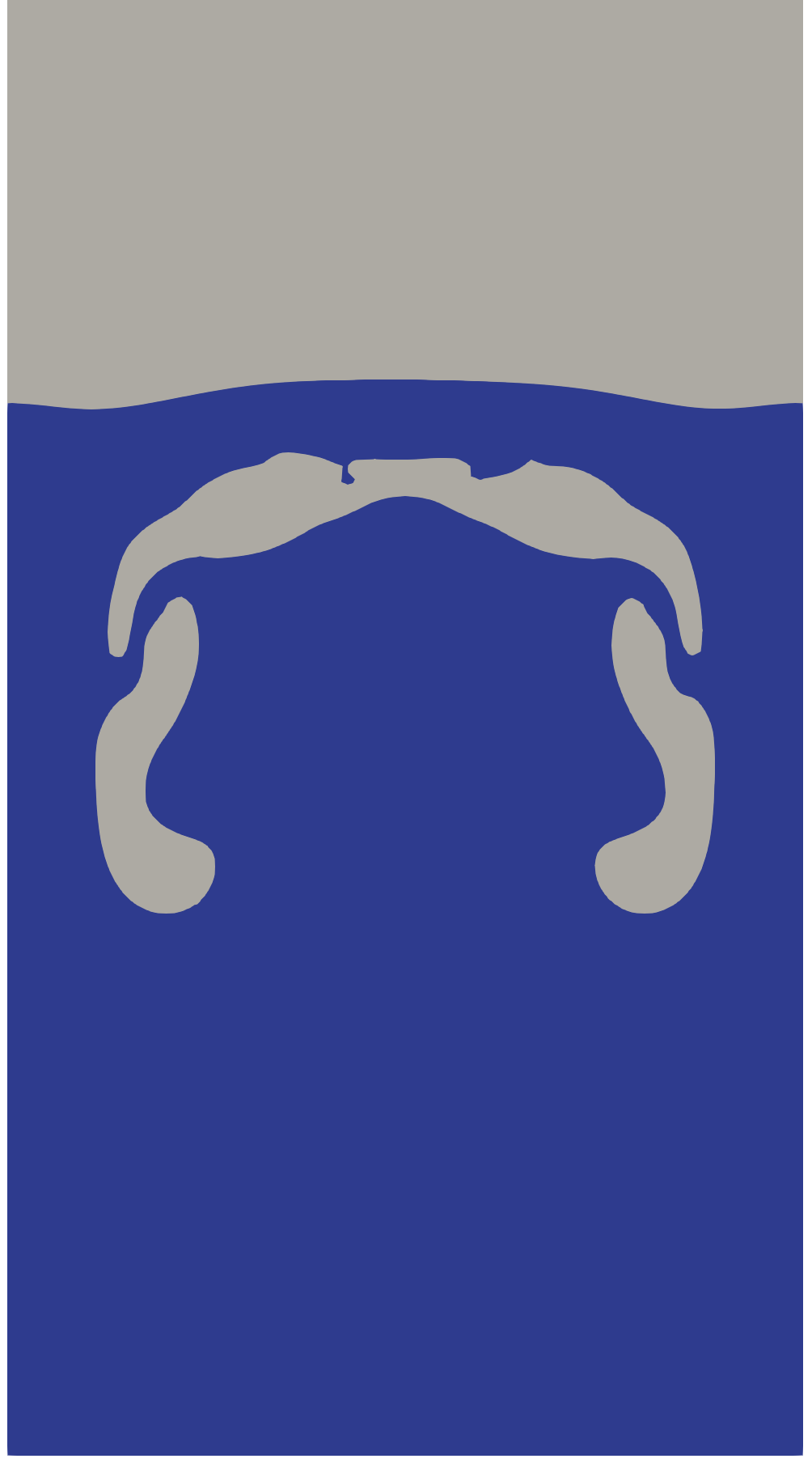}
        \caption{$t=2.2$}
    \end{subfigure}
    \hfill
    \begin{subfigure}[b]{0.24\textwidth}  
        \centering 
        \includegraphics[width=\textwidth]{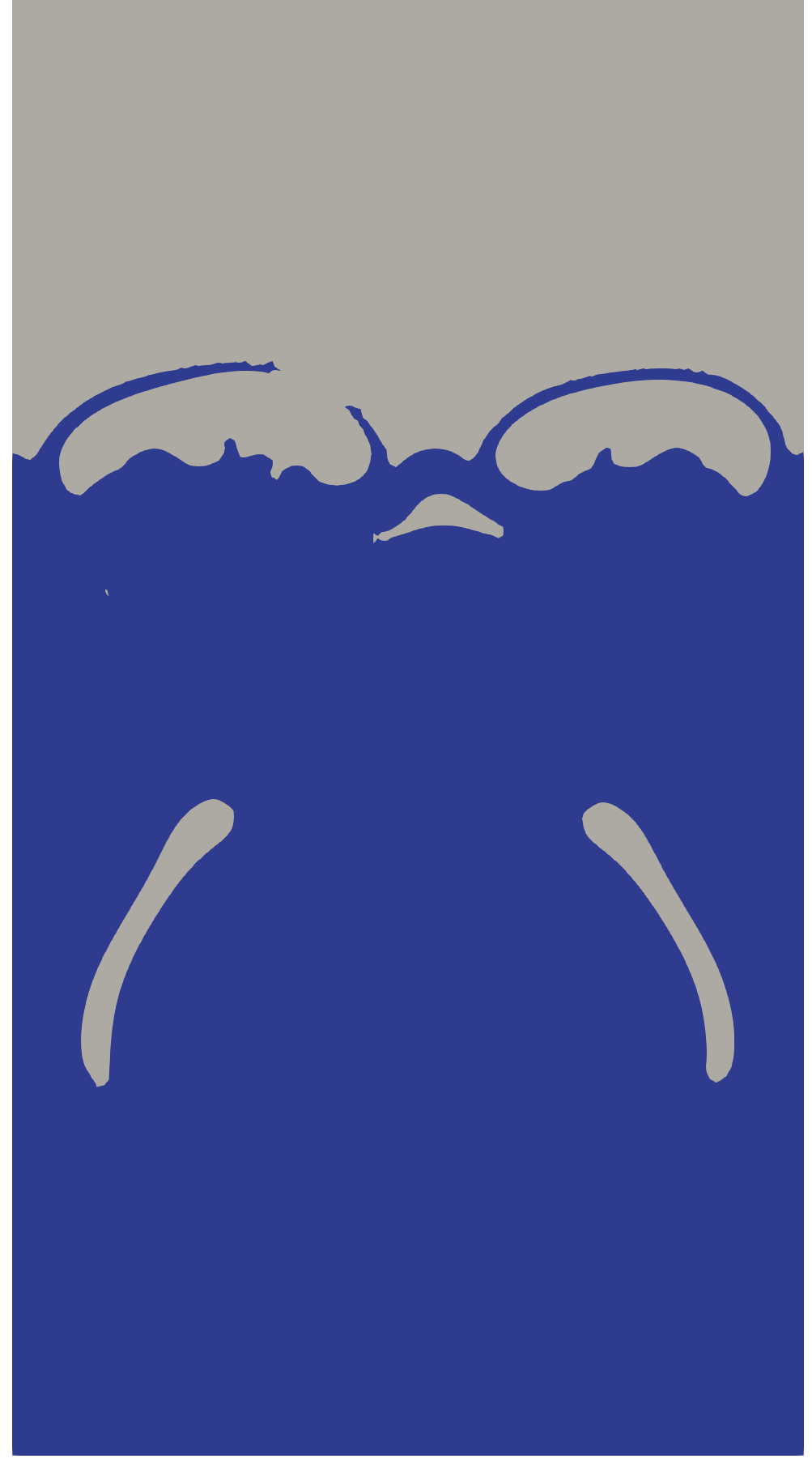}
        \caption{$t=3.85$}
    \end{subfigure}
\end{figure}

\section{Conclusion}
In this paper we have presented an approach to simulate two-phase flows by modeling the interface in a sharp and low cost way using X-Mesh. The representation of the interface directly in the mesh allows it to be sharp. It is able to represent discontinuities in the derivatives of velocities at the interface like the ALE methods. The update of the interface position by means of a level set function allows to easily take into account the changes of the fluid phase topology. The front relay allows the good representation of the large displacements of the interface by carrying out only a local deformation, close to the interface, of the mesh. The presented deformation algorithm allows to keep the original mesh and thus to preserve its topology. 

The surface tension model is exactly sharp thanks to the positioning of the nodes on the interface and the imposition of the pressure jump as in the ghost fluid method.  Thanks to the calculation of the curvature of the interface via the circumscribed circle of 3 successive nodes, the parasitic currents have been reduced to the order of machine precision. Finally, the sequential algorithm allows to couple the different steps of the simulation in a robust way despite the numerous changes of the fluid phase topology. 

However, mass variations have been observed in the simulations. Although the level set advection scheme is exactly conservative with respect to the global integral of the level set field, this integral does not represent anything physical. The integral of the level set can be exactly conserved and still we could observe large mass variations. This is inherent in the level set approach and that's why we want to improve this in future work by changing the interface representation. Let's also note that the PSPG stabilisation introduces an error mass conservation equation and the resulting velocity field is not exactly divergence free.

\newpage
\section*{Acknowledgements}
This project has received funding from the European Research Council (ERC) under the European Union's Horizon research and innovation programme (Grant agreement No. 101 071 255)

The authors thank Michel Henry for maintaining the features developed in MigFlow and Nicolas Chevaugeon for his help in implementing the fastmarching algorithm
%%Vancouver style references.
\bibliographystyle{model1-num-names}
\bibliography{refs}

\end{document}